\documentclass[prx,superscriptaddress,reprint,showpacs,longbibliography]{revtex4-1}
\usepackage{color}
\usepackage{graphicx,textcomp,amssymb,amsmath,dcolumn,hyperref}
\usepackage{siunitx}
\usepackage{comment}

\begin{document}

\title{Tunable valley splitting and bipolar operation in graphene quantum dots}

\author{C.~Tong}
\email{ctong@phys.ethz.ch}
\author{R.~Garreis}
\affiliation{Solid State Physics Laboratory, ETH Zurich, CH-8093 Zurich, Switzerland}
\author{A.~Knothe}
\affiliation{National Graphene Institute, University of Manchester, Manchester M13 9PL, United Kingdom}
\author{M.~Eich}
\author{A.~Sacchi}

\affiliation{Solid State Physics Laboratory, ETH Zurich, CH-8093 Zurich, Switzerland}

\author{K.~Watanabe}
\author{T.~Taniguchi}
\affiliation{National Institute for Material Science, 1-1 Namiki, Tsukuba 305-0044, Japan}

\author{V.~Fal'ko}
\affiliation{National Graphene Institute, University of Manchester, Manchester M13 9PL, United Kingdom}
\affiliation{Henry Royce Institute for Advanced Materials, M13 9PL, Manchester, UK}

\author{T.~Ihn}
\author{K.~Ensslin}
\author{A.~Kurzmann}
\affiliation{Solid State Physics Laboratory, ETH Zurich, CH-8093 Zurich, Switzerland}

\date{\today}

\begin{abstract}
 
Quantum states in graphene are four-fold degenerate: two fold in spins, and two fold in valleys. Both degrees of freedom can be utilized for qubit preparations. In our bilayer graphene quantum dots, we demonstrate that the valley g-factor $g_\mathrm{v}$, defined analogously as the spin g-factor $g_\mathrm{s}$ for valley splitting in perpendicular magnetic field, is tunable by over a factor of 4 from 20 to 90. We find that larger $g_\mathrm{v}$ results from larger electronic dot sizes, determined from the charging energy. This control is achieved by adjusting voltages on merely two gates, which also allows for tuning of the dot-lead tunnel coupling. On our versatile device, bipolar operation, charging our quantum dot with charge carriers of the same or the opposite polarity as the leads, can be performed. Dots of both polarity are tunable to the first charge carrier by action of the plunger gate, such that the transition from an electron to a hole dot can be observed. By adding more gates, this system can easily be extended to host double dots.

\end{abstract}
	
\maketitle
 
\section{Introduction}

A local minimum or maximum in the electronic band structure of a solid is called a valley. These valleys are relevant for the electronic properties of a number of materials, such as Si \cite{Sze69}, AlAs \cite{Shayegan06}, graphene and other 2D materials \cite{Schaibley16}. Energies of valley states can be tuned by strain, a feature utilized by the semiconductor industry to increase carrier mobility \cite{Thompson04}. For the purpose of constructing qubits, the valley degree of freedom can be treated similarly to orbital and spin degrees of freedom \cite{Rohling12}. So far, few experimental reports exist in this direction \cite{Penthorn19}.

Theories for spin qubits \cite{Loss98} foresee local tunability of the electronic g-factor $g_\mathrm{s}$, relevant for the Zeeman splitting of energy levels in a magnetic field. Such tunability enables different qubits to be selectively addressed, with the magnetic field as their common control. This has been achieved in several experiments \cite{Salis01,Kato03,Nitta03,Bjork05} with changes in $g_\mathrm{s}$ of a few percent. For valley qubits, one would similarly like to tune the valley g-factor $g_\mathrm{v}$, which characterizes the valley splitting that is linear in low magnetic fields. Such tunability is however thus far limited \cite{hollmann2020large}. 

Graphene quantum dots (QDs) have been proposed as spin qubits with long spin coherence times \cite{Trauzettel2007spinqubit} due to weak spin-orbit interaction and hyperfine coupling. As an alternative, the two valleys of graphene offer a pseudo-spin degree of freedom, making graphene QDs also amenable for realizing valley qubits. Experiments on graphene quantum point contacts have already demonstrated that $g_\mathrm{v}$, arising from orbital magnetic moments related to the Berry curvature, can be tuned by gate voltages by a factor of three \cite{Lee20}.

Here we demonstrate that, $g_\mathrm{v}$ in a few-carrier bilayer graphene (BLG) QD can be {\em in situ} tuned via purely electrostatic means in a controlled and systematic fashion by a factor of 4.5 from $g_\mathrm{v}=20$ to 90. We find experimentally that the tunability of $g_\mathrm{v}$ is related to the electronic size of our QD, which scales with its charging energy. Theory relates the value of $g_\mathrm{v}$ and its tunability to the same Berry curvature effects found relevant in quantum point contacts \cite{angelika2018minivalley,angelika2020quartet}. We show that the results of detailed modeling are in qualitative and semi-quantitative agreement with the experiment. Tuning of the charging energy and hence of $g_\mathrm{v}$ is achieved with a carefully designed and yet straightforward three-gate geometry, which provides us also with control over the tunnel coupling between the QD and its leads, as well as over the polarity of charge carriers in the dot. Such control allows for bipolar operation of the QD, enabling us to observe the transition from an electron dot to a hole dot through the first carrier states. This device geometry provides a straightforward extension to double quantum dots.

\section{Tunable valley splitting}

\begin{figure*}
	\includegraphics{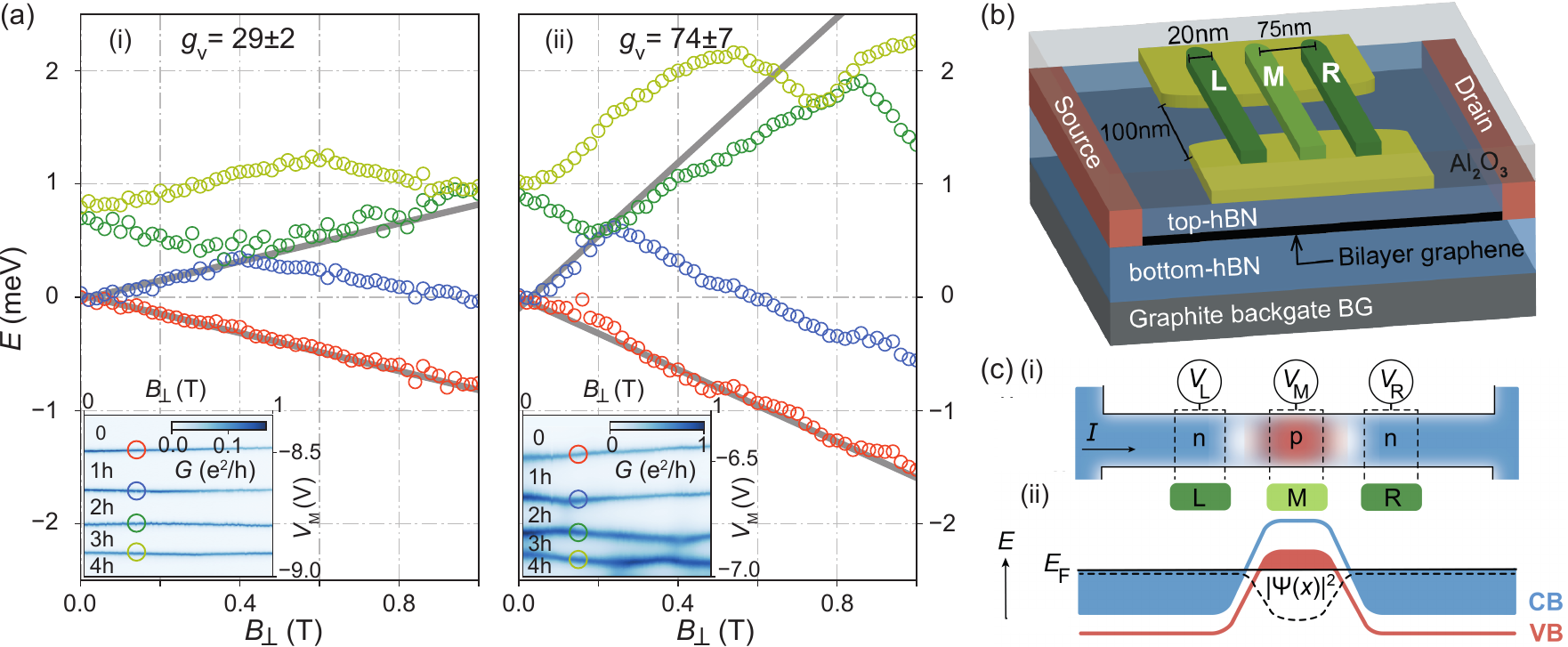}
	\caption{(a) Single particle energy level spectrum of a QD at (i) $V_\text{BG}=\SI{4.8}{V}$, $V_\mathrm{SG}=\SI{-3.44}{V}$, $V_\text{L,R}=\SI{0.0}{V}$ and (ii) $V_\text{BG}=\SI{3.0}{V}$, $V_\mathrm{SG}=\SI{-2.45}{V}$, $V_\text{L,R}=\SI{4.0}{V}$ in perpendicular magnetic field $B_\mathrm{\perp}$, extracted from conductance maps shown accordingly as insets. Extraction of energy levels is performed similarly as in Ref.~\cite{Mariusprx,annikaexcitedstates} (see \ref{subsection:extraction} for more details), such that resonances are vertically shifted to touch their nearest neighbors at points with the smallest separations. Levels extracted from the $1^\text{st}$, $2^\text{nd}$, $3^\text{rd}$ and $4^\text{th}$ hole-state Coulomb resonance peaks are plotted in red, blue, green and yellow, respectively. Gray lines indicate indicate the linear valley splitting. (b) 3D illustration of the van der Waals hetero-structure device with layers labeled and gates shown: split gates in yellow, and finger gates L, M and R in green. Edge contacts (orange) are fabricated at two ends of the channel. (c) Schematics of a hole dot formed in a $n$-type channel, with no action from gates L and R. (i) Top-view of the channel where split gates are outlined in solid and finger gates in dashed lines. Red, blue and white indicate $n$-type, $p$-type, and gapped regions, respectively. (ii) Corresponding conduction (blue) and valence (red) band edge structure along the channel. The extent of the probability distribution of the QD hole state $|\Psi(x)|^{2}$ is sketched by a dashed line.}
	\label{figbegin}
\end{figure*}

The essence of our main result is highlighted in Fig.~\ref{figbegin}(a), which is accessible without a detailed description of our BLG QDs. The single particle level spectra for the QDs show energy splitting in a perpendicular magnetic field for the first four hole states lowest in energy. Indicated by the gray lines, this splitting is linear at low fields, and corresponds to the energy splitting between neighboring levels belonging to the $K^-$ and $K^+$ valleys. The valley g-factor $g_\mathrm{v}$ is extracted from $\Delta E_{K^-,K^+}= g_\mathrm{v}\mu_\mathrm{B}B_\perp$. The splitting in (a,ii) is evidently much larger than that in (a,i) (quantitatively, the $g_\mathrm{v}$ values differ by a factor of 2.5), where (i) and (ii) are taken for QDs formed with the same gates but at two different gate voltage configurations. We attribute this salient difference in $g_\mathrm{v}$ to small (i) and large (ii) electronic dot sizes. We now start with a detailed description of how these results are obtained, and later demonstrate an even larger tuning range for $g_\mathrm{v}$ by a factor of 4.5.

A schematic illustration of the device is shown in Fig.~\ref{figbegin}(b) (see \ref{subsection:sample} for details on sample fabrication). With the mutual action of a graphite back-gate voltage $V_\mathrm{BG}$, and a top-gate voltage, we can alter both the Fermi energy $E_\mathrm{F}$, and the size of the BLG band-gap $\Delta_\text{gap}$, below the respective top gates. Top gates here can be either the split gates (yellow, separated by \SI{100}{nm}) or the finger gates (green, \SI{20}{nm} in width and separated from each other by \SI{75}{nm} center to center). The band-gap $\Delta_\text{gap}$ scales with the strength of the displacement field perpendicular to the BLG sheet \cite{Ohta2006BLGband,mccann2006BLGband,Oostinga2008BLG}. Here we operate at positive $V_\mathrm{BG}$, inducing $n$-type charge carriers in the bulk BLG regions.

The situation is depicted in Fig.~\ref{figbegin}(c,i). For a fixed positive $V_\text{BG}$, negative split-gate voltages $V_\mathrm{SG}$ open a band-gap underneath the split gates and tune $E_\mathrm{F}$ into this gap, confining $I$ to be along the $n$-type conducting channel as indicated by the arrow. We then operate our finger gates at fixed $V_\text{SG}$. Negative voltages $V_\mathrm{M}$ on gate M locally generate a $p$-type island in the channel. Thus a $p$-type QD is formed with naturally arising $p$-$n$ junctions as tunnel barriers to the $n$-type leads, with gate M as its plunger gate. In this regime, we observe signatures of Coulomb blockade in the quantum dot, with sharp conductance resonances as a function of $V_\text{M}$ and fully suppressed current in-between. The corresponding band edge structure along the channel axis is depicted in (c,ii). The tunnel barriers are tuned by voltages $V_\text{L,R}$ applied to the barrier gates L and R (dark green), thus allowing for control of the electronic size of this quantum dot, and also its tunnel coupling to the leads.

\begin{figure}
	\includegraphics[width=8.6cm]{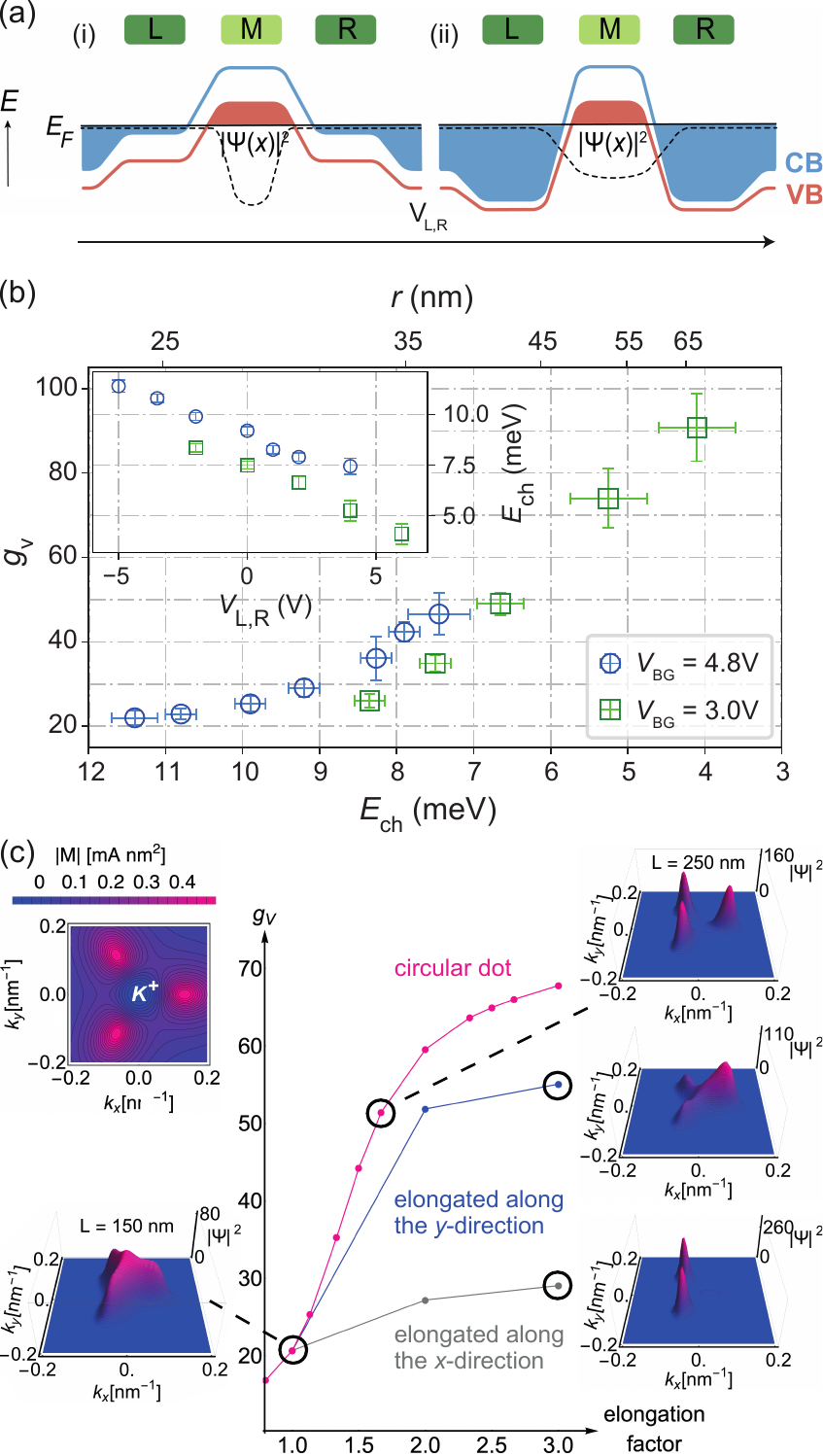}
	\caption{(a) Schematic representation of conduction (blue) and valence (red) band edge structures along the channel axis, controlled by finger gates L, M, and R on top. Extent of the probability distribution of the QD hole state $|\Psi(x)|^{2}$ is sketched in dashed lines. Barrier voltages $V_\mathrm{L,R}$ becomes more positive from left to right. (b) Extracted $g_\mathrm{v}$ as a function of decreasing charging energies (i.e. increasing electronic dot radius). The inset shows corresponding $V_\mathrm{L,R}$ configurations for these charging energies. Blue circles and green squares are measured at $V_\text{BG}=\SI{4.8}{V}$, $V_\mathrm{SG}=\SI{-3.44}{V}$ and $V_\text{BG}=\SI{3.0}{V}$, $V_\mathrm{SG}=\SI{-2.45}{V}$ respectively. (c) Calculated momentum space wave functions and valley g-factors for dots of different sizes and shapes. For an elongation factor of 1, the dot is circularly symmetric with a radius of \SI{150}{nm}. When elongating in both directions (magenta line), or along the $x$ or $y$-axis yielding elliptical dots (gray and blue line, respectively), the valley g-factor increases when the momentum space wave functions squeeze into the mini-valleys of the BLG dispersion. Inset (top left) shows the orbital magnetic moment, $M$, of the unperturbed BLG's first conduction band around the $K^+$ valley and for a gap of $\Delta_\text{gap}=\SI{42}{meV}$.}
	\label{figpngV} 
\end{figure}

We find experimentally that the dot-lead tunnel coupling can be tuned by $V_\mathrm{L,R}$, where positive $V_\mathrm{L,R}$ leads to a more strongly coupled dot, and negative $V_\mathrm{L,R}$ to a more decoupled one. The peak conductance $G_\text{max}$ can be tuned by three orders of magnitude ($\SI{}{10^{-3} e^{\mathrm{2}}/h} - \SI{}{e^{\mathrm{2}}/h}$), and tunneling rates $\Gamma$ by a factor of 30 ($\SI{10}{GHz}-\SI{300}{GHz}$). Details about measurements and extraction of these numbers are presented in \ref{subsection:coupling}.

Changes in $V_\text{L,R}$ also affect the quantum dot size. As an indicator of its size, the dot's charging energy $E_\text{ch}$ is extracted from Coulomb diamond measurements of the first hole. The inset of Fig.~\ref{figpngV}(b) shows $E_\text{ch}$ linearly decreasing by a factor of four (representing an increase in the QD size) for two different back gate voltages as $V_{L,R}$ is varied from negative to positive over the available tuning range. Reliable extraction of smaller $E_\mathrm{ch}$ is limited by the smearing of the resonance peaks and instability of the dots due to their stronger coupling to the leads. Dots formed at lower $V_\text{BG}$ and thus at a weaker displacement field have systematically smaller $E_\text{ch}$ and thus larger sizes, due to the smaller BLG gap $\Delta_\text{gap}$.

To obtain a rough estimate for the quantum dot size, the charging energy $E_\mathrm{ch}$ is converted into radius, assuming the QD to be a circular disk-like capacitor of radius $r$, embedded deep in a mixture of insulating hBN and amorphous Al$_2$O$_3$, with self-capacitance $C=8\epsilon_{\text{r}}\epsilon_0r$ \cite{Thomasbook}. A decrease in $E_\mathrm{ch}$ from $\SI{11.4}{meV}$ to $\SI{4.1}{meV}$ therefore corresponds to an increase in dot size from a radius of $\SI{24}{nm}$ to $\SI{67}{nm}$. In reality, our QDs vary in aspect ratios: As $V_\mathrm{BG}$ and $V_\mathrm{SG}$ are kept the same, the extent of the QD across the channel remains roughly constant while $V_\mathrm{L,R}$ expands and contracts the dot only along the channel.

Our high tunability is enabled by the narrow finger gates with small separations in-between. The former ensures formation of small enough structures, whereas the latter allows for sufficient control over the bands not only directly underneath but also in-between the finger gates by stray fields. 

The band-edge structure schematics shown in Fig.~\ref{figpngV}(a) illustrate the above experimental observations of our QD under more negative (i) and more positive (ii) $V_\mathrm{L,R}$. Similar to Fig.~\ref{figbegin}(c,ii), $V_\mathrm{BG}$ and $V_\mathrm{SG}$ are kept constant, and a negative $V_\mathrm{M}$ has lifted the bands up in energy to form a $p$-type dot, with $p$-$n$ junctions as tunnel barriers. When more negative barrier voltages $V_\mathrm{L,R}$ are then applied, the bands are raised in energy underneath gates L and R; $\Delta_\text{gap}$ is also increased due to the larger displacement field. As illustrated in (i), these changes effectively induce shallower, and more importantly, wider $p$-$n$ junctions along the channel, such that the tunnel barriers become more opaque. On the other hand, when $V_\mathrm{L,R}$ is more positive in (ii), bands underneath L and R are lowered in energy and $\Delta_\text{gap}$ is decreased. The $p$-$n$ junctions become steeper and narrower, forming more transparent tunnel barriers. As a result, the dot-lead tunnel coupling increases in case (ii) [decreases in case (i)] at more positive (more negative) $V_\mathrm{L,R}$, giving rise to broader (narrower) Coulomb resonance peaks with higher (lower) peak conductance. 

Tunable dot size with $V_\mathrm{L,R}$ then follows straightforwardly. With more opaque (transparent) barriers from more negative (positive) $V_\mathrm{L,R}$, charges are more (less) confined in the $p$-type dot. Sketched with dashes in Fig.~\ref{figpngV}(a), the extent of the probability distribution $|\Psi(x)|^{2}$ of the QD therefore spreads less (more) into the leads, forming an effectively smaller (larger) dot. According to experimental data, this effect dominates over the {\em a priori} anticipation that the cross-capacitance between gates L and R with gate M will decrease (increase) the dot size with more positive (negative) $V_\mathrm{L,R}$.

\begin{figure*}
	\includegraphics[width=18.6cm]{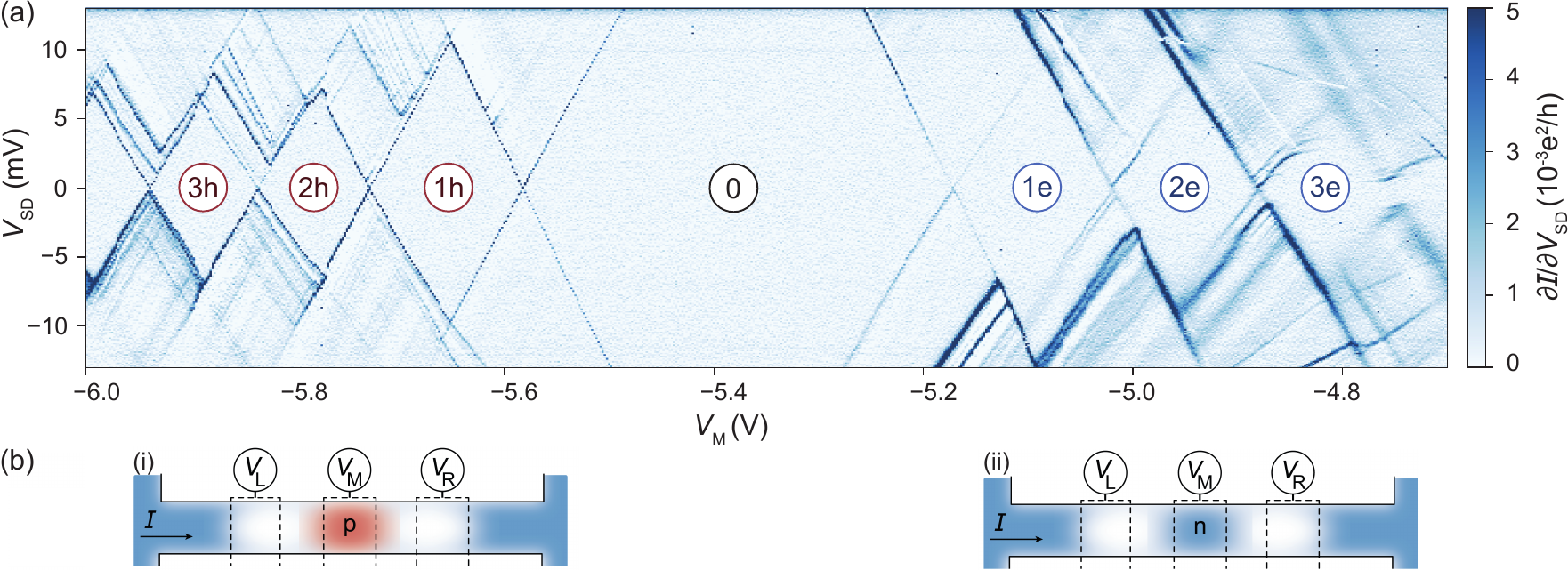}
	\caption{(a) Coulomb diamonds showing the bipolar operation of our QD. QD carrier occupancy is marked on the corresponding diamonds. (b) Schematic top-view of the channel where split gates are outlined in solid and finger gates in dashed lines, when (i) a hole dot and (ii) an electron dot is formed in a $n$-type channel. Red, blue and white indicate $p$-type, $n$-type, and gapped regions, respectively. }
	\label{figehsing}
\end{figure*}

With the understanding of the tuning of our quantum dot size, we are now in a position to explain how we obtain our tunable valley g-factors. At fixed $V_\text{BG}$ and $V_\text{SG}$, we measure Coulomb resonances for the first four holes as a function of the plunger gate voltage $V_\text{M}$ in perpendicular magnetic field $B_\perp$, for QDs formed at a set of different $V_\text{L,R}$. Examples of such measurements are shown as insets of Fig.~\ref{figbegin}(a), with carrier occupancy labeled. We convert the $V_\text{M}$-axis into energy, using lever arms determined from Coulomb diamond measurements at corresponding gate voltage configurations. We then extract the single particle level spectrum, by subtracting from the conductance resonance positions (i.e. the energy levels) a magnetic field independent charging energy (see \ref{subsection:extraction} for details). Examples of the resulting spectra are shown in Fig.~\ref{figbegin}(a) for dots with charging energies $E_\mathrm{ch}=\SI{9.2}{meV}$ for (i) and $E_\mathrm{ch}=\SI{4.8}{meV}$ for (ii). Indicated by the gray lines, at sufficiently low magnetic fields, the lowest two energy levels, being states of the $K^-$ and $K^+$ valleys, split linearly in energy $\Delta E_{K^-,K^+}=g_\mathrm{v}\mu_\text{B}B$. From $\Delta E_{K^-,K^+}$ for the first and the second level, before the crossing with the third level occurs, we determine the valley g-factor $g_\mathrm{v}$.

The extracted values of $g_\mathrm{v}$ are plotted against the corresponding charging energies $E_\mathrm{ch}$ in Fig.~\ref{figpngV}(b). Evidently, dots with lower charging energy (i.e. larger electronic size) have a systematically larger $g_\mathrm{v}$. The tuning range is as large as a factor of 4.5 varying from $g_\mathrm{v}=90$ at $E_\mathrm{ch}=\SI{4.1}{meV}$ ($r=\SI{67}{nm}$) to $g_\mathrm{v}=20$ at $E_\mathrm{ch}=\SI{11.4}{meV}$ ($r=\SI{24}{nm}$). These results encompass values obtained in previous measurements of $g_\mathrm{v}$ in BLG QDs \cite{Mariusprx,annikaexcitedstates,aachenehcrossoover}. 

We present calculations of the dot's electronic structure in support of our experimental observations. We describe electrostatic confinement by a smooth potential \cite{angelika2020quartet} where we elongate the radius in both directions (the $x$- or the $y$- axis) to model rotationally symmetric (elliptic) QDs. See \ref{section:Theory} for details of the model. The valley splitting $\Delta E_{K^{-},K^{+}}=g_\mathrm{v}\mu_\mathrm{B}B$ is expressed in terms of the valley g-factor $g_\mathrm{v}$, which arises from  the orbital magnetic moment $M$ in BLG k-space, originating from the material's non-trivial Berry curvature. In BLG the band edges at the three-fold symmetrical mini-valleys possess the largest $M$ [top-left inset to Fig.~\ref{figpngV}(c)]. For sufficiently large dots, increasing the dot's radius along one or both axes in real space squeezes the wave-functions into the mini-valleys in momentum space, which then dominates the amount of $M$ being picked up. Hence, as shown in Fig.~\ref{figpngV}(c), $g_\mathrm{v}$ increases with increasing dot sizes as the contributions from the mini-valleys become dominant. For non-circular dots (closer to our experimental operations as discussed before), $g_\mathrm{v}$ depends on the elongation direction. Our experimental observation corresponds well with predictions in the large-dot regime (see \ref{section:Theory} for more data as a function of the dot size and a discussion of the small-dot regime). 

\section{bipolar operation}
\label{section:bipolar}

The exquisite device control achieved with gates L, M and R not only enables us to tune the valley splitting, but also allows us to explore QDs with carriers of both polarities at will. In the previous section we considered a $p$-type dot in an $n$-type channel, with $p$-$n$ junctions as its tunnel barriers. Now we consider the situation where the dot is defined by gapped regions underneath the barrier gates L and R, with $E_\text{F}$ tuned into the gap. In this configuration, bipolar operation of the quantum dot is possible and the transition from a quantum dot occupied with the first electron to one occupied with the first hole can be observed by tuning gate M only. 

To this end, we operate the quantum dot at $V_\mathrm{BG}=\SI{3.7}{V}$ and $V_\mathrm{SG}=\SI{-2.92}{V}$, forming again a $n$-type channel between the split gates, and insulating regions beneath them. Negative $V_\text{L,R}$ depletes the electrons locally and tunes $E_\text{F}$ into the band gap below gates L and R. A conductive region below gate M is thereby isolated from the leads, which becomes the QD with charge carriers trapped inside. Tuning $V_\text{M}$ to sufficiently negative
values enables us to change the type of charge carriers in the dot from electrons to holes, thereby facilitating bipolar operation of the dot. Due to the close separation of the finger gates, finite cross-talk exists between the plunger gate M and the barrier gates L and R. In order to mitigate the cross-talk and to keep the tunnel coupling approximately constant, we linearly correct $V_\mathrm{L,R}$ when $V_\text{M}$ is swept, instead of operating at constant $V_\mathrm{L,R}$. Details are shown in \ref{subsection:VbVp}.

The transition between the electron and the hole dot can be seen from the finite bias measurement shown in Fig.~\ref{figehsing}(a), taken at the dashed line on Fig.~\ref{figVbVp}. Carrier occupancy is labeled in the corresponding diamonds. The transition of polarity from a $n$-type to a $p$-type quantum dot is identified as the largest diamond-shaped region (labeled 0) extending from $V_\text{M}=\SI{-5.2}{V}$ to \SI{-5.6}{V}. Well-defined Coulomb diamonds with clearly visible excited states are shown for both electrons and holes for the first three carriers. We find similar lever arms and charging energies for carriers on both sides of the gap, indicating a similar geometry for the dots of the two polarities. The fine resolution of a rich group of excited states \cite{annikaexcitedstates} is a strong indication of the cleanliness and the high quality of our QD. From the size of the diamond labeled with carrier occupancy 0, we extract $\SI{30}{meV}$ as the energy separation between the first electron and the first hole state, providing a lower bound estimate of the band gap in the dot. 

A schematic representation of the dots with both polarities is sketched correspondingly below the measurements in Fig.~\ref{figehsing}(b). In the previous section, adjusting $V_\mathrm{L,R}$ over a large range tuned the width of the $p$-$n$ junction. Now $V_\mathrm{L,R}$ changes the size of the band gap and the position of $E_\text{F}$ in the gap, and thus tunes the height of the tunnel barriers. The range of $V_\mathrm{L,R}$ suitable for the formation of such a dot is smaller compared to the range in effect in the previous section, as we are limited by the requirement that regions underneath gates L, R need to remain in the gap. The dot-lead coupling strength for such a dot is tunable by $V_\mathrm{L,R}$, which has been demonstrated in previous work \cite{mariusnanolett}, though there the first carrier state was not observed due to the large separation between barrier and plunger gates. In our case, a negative $V_\mathrm{M}$ depletes the QD into the few-electron regime [sketched in (b,ii)], and eventually depletes it after removal of the last electron. More negative $V_\text{M}$ then adds single holes one-by-one into the now $p$-type dot [sketched in (b,i)]. During the preparation of this manuscript, we became aware of similar measurements on this bipolar operation, with a different device geometry \cite{aachenehcrossoover}. 

\begin{figure}
	\includegraphics[width=8.6cm]{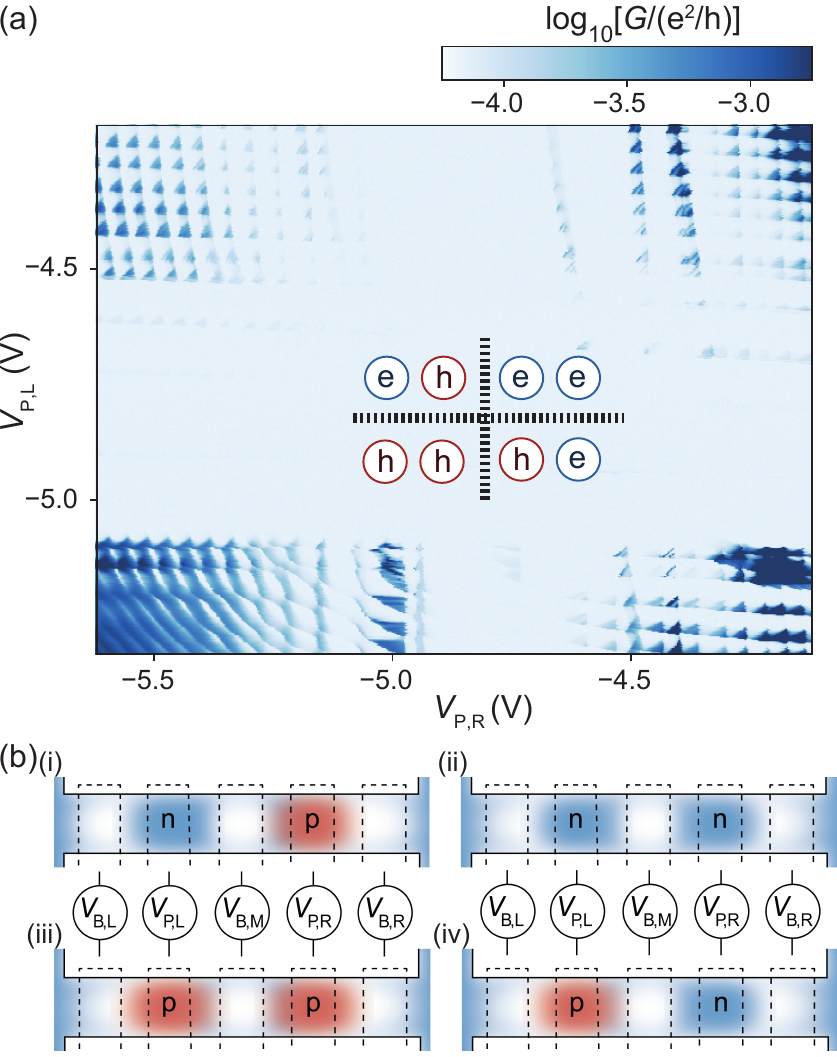}
	\caption{(a) Charge stability diagram of a double dot system formed using five finger gates. Separated by the gap of low conductance, carrier types of the four quadrants are labeled in the middle of the conductance map. (b) Schematic top-view of the channel, where split gates are outlined in solid and finger gates in dashed lines, with $n$-type, $p$-type, and gapped regions colored blue, red, and white, respectively. (i)-(iv) corresponds to the four quadrants in (a).}
	\label{figdd}
\end{figure}

Within the same device, but using five finger gates instead of three, we are also able to form bipolar double dots. Repeating the tuning procedure discussed above, and choosing appropriate barrier voltages $V_\mathrm{B}$, both dots are tunable in carrier occupation to the last electron/hole. Moreover, just like for the single dot case, control over the dot-lead and also the dot-dot tunnel coupling is possible. Figure~\ref{figdd}(a) presents the charge stability diagram of this bipolar double dot taken at barrier gate voltages $V_\mathrm{B,L}=\SI{-5.5}{V}$, $V_\mathrm{B,M}=\SI{-5.7}{V}$ and $V_\mathrm{B,R}=\SI{-5.5}{V}$, as a function of the two plunger gate voltages $V_\mathrm{P,L}$ for the left and $V_\mathrm{P,R}$ for the right dot. We perform this measurement at a finite source-drain bias voltage of \SI{1}{mV} to enhance the visibility of the triple points.

Four different combinations of double dot polarity are observed, with finite-bias triangles visible in all four quadrants, separated by the gap of low conductance. Labeled in the center of the conductance map in Fig.~\ref{figdd}(a), with corresponding schematics shown below in (b), we form (i) electron-hole (ii) electron-electron (iii) hole-hole and (iv) hole electron double dots, respectively, with $n$-type leads. 

\section{Conclusion}

We have presented a BLG device, allowing for {\em in situ} formation of bipolar quantum dots with tunable coupling strengths, electronic dot sizes and valley g-factors. The valley g-factor can be tuned by a factor of 4.5 from $g_\mathrm{v}=20$ to 90 with increasing electronic dot sizes. This dependence is brought about by the increasing influence of the high magnetic moments of the orbital mini-valleys, in qualitative agreement with theoretical calculations. The system can be straight-forwardly extended to host double dots, and potentially other multi-dot systems, by simply adding more gates. Our versatile BLG QD with widely tunable valley g-factor paves the way for promising valley qubits in graphene. 

\section*{acknowledgments}

We thank P. Märki and T. Bähler as well as the FIRST staff for their technical support. We also acknowledge financial support from the European Graphene Flagship, the Swiss National Science Foundation via NCCR Quantum Science and Technology, the EU Spin-Nano RTN network, and the European Union’s Horizon 2020 research and innovation programme under the Marie Skłodowska-Curie Grant Agreement No. 766025. Growth of hexagonal boron nitride crystals was supported by the Elemental Strategy Initiative conducted by the MEXT, Japan and JSPS KAKENHI Grant No. JP15K21722. We acknowledge funding from the Core3 European Graphene Flagship Project, the European Quantum Technology Project 2D-SIPC, the ERC Synergy Grant Hetero2D, and EPSRC grants EP/S030719/1 and EP/N010345/1.

\setcounter{section}{0} 
\renewcommand\thesection{Appendix~\Alph{section}} 

\section{Materials and Methods}
\renewcommand\thesubsection{\arabic{subsection}} 
\subsection{Sample fabrication}
\label{subsection:sample}

The device is fabricated as described in \cite{Hiske2018electrostatically, Mariusprx}, and schematically depicted in Fig.~\ref{figbegin}(b). Built with the dry-transfer technique \cite{wang2013drytransferedge}, the van der Waals hetero-structure stack lies on a silicon chip with \SI{280}{nm} surface SiO$_2$. The stack consists of a bottom graphite back gate [dark gray in Fig.~\ref{figbegin}(b)], and on top of it a BLG flake (black) encapsulated in \SI{38}{nm} thick bottom and \SI{20}{nm} thick top hBN flakes (blue). Ohmic edge contacts (red) with Cr and Au of \SI{10} and \SI{60}{nm} thickness, respectively, are evaporated after etching through the top hBN flake with reactive ion etching. A pair of \SI{5}{n}m thick Cr, \SI{20}{nm} thick Au split gates (yellow), separated by \SI{100}{nm} are deposited on top, defining a $\SI{1}{\mu m}$ long channel. Separated by a layer of \SI{30}{nm} thick amorphous Al$_2$O$_3$ (light gray) grown by atomic layer deposition, finger gates (green) labeled L, M and R of \SI{20}{nm} in width, and \SI{5}{nm} Cr and \SI{20}{nm} Au in thickness, lay across the channel defined by the split gates. Nearest neighbor finger gates are separated by $\SI{75}{nm}$ from center to center.

\subsection{Measurement condition}
\label{subsection:measure}
We perform AC (standard lock-in techniques) and DC measurements, in He$^{3}$/ He$^4$ dilution refrigerators, at electronic temperatures of around $\SI{100}{mK}$.

\subsection{Extracting single particle level spectra}
\label{subsection:extraction}

Similar to \cite{Mariusprx, annikaexcitedstates}, we extract single particle level spectra from Coulomb resonance peak positions by following their evolution in perpendicular magnetic field $B_\mathrm{\perp}$ [insets in Fig.~\ref{figbegin}(a)]. Lorentzian functions are fitted to each peak, obtaining peak positions in plunger gate voltage $V_\mathrm{M}$. These gate voltage values are then converted into energies, using lever arms $\alpha$'s extracted from Coulomb diamond measurements taken at corresponding gate voltage configurations ($V_\mathrm{M}$, $V_\mathrm{L,R}$, $V_\mathrm{BG}$, and $V_\mathrm{SG}$). The first and the second level are converted with lever arms $\alpha_\mathrm1$ extracted from diamonds of the first carrier state, and $\alpha_\mathrm2$ extracted from the second carrier state, respectively. The third and the fourth level are converted using an average of $\alpha_\mathrm1$ and $\alpha_\mathrm2$. In practice, $\alpha_\mathrm1$ and $\alpha_\mathrm2$ differ from each other within the error range. 

The energy difference between the nearest neighboring levels is the corresponding addition energy, which is the energy required to add a carrier to the QD. This energy consists of not only the single particle energy difference, but also the charging energy stemming from the Coulomb interaction, which we assume to be independent of magnetic field. When two single particle levels are degenerate, the addition energy comprises of only the charging energy. We therefore assume the minimal difference between nearest neighboring levels (occurring at anti-crossings or at zero magnetic field) to be exactly the charging energy. For clarity, we subtract charging energies from these levels, and as a result shift levels vertically in energy such that nearest neighboring levels touch at the point where they are degenerate. We also shift the origin of the energy axis to be the energy of the first level at zero magnetic field. Examples of the resulting single particle energy spectra are presented in Fig.~\ref{figbegin}(a). This way of extraction sees a lifting of spin degeneracy at zero magnetic field. 

\section{Supporting Data}
\label{section:Support}

\subsection{Tunable tunnel coupling}
\label{subsection:coupling}
\begin{figure}
	\includegraphics[width=8.6cm]{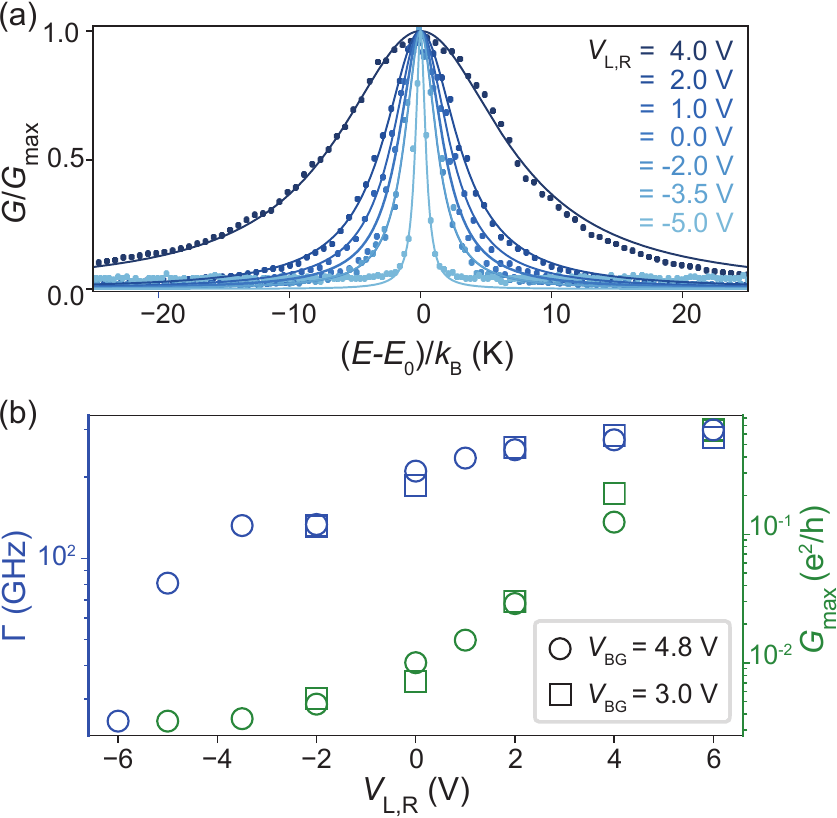}
	\caption{(a) Coulomb resonances of the first hole at $V_\mathrm{BG}=\SI{4.8}{V}$ and $V_\mathrm{SG}=\SI{-3.44}{V}$, plotted in energies converted into units of temperature, for varying barrier gate voltages $V_\mathrm{L,R}$, scaled by their peak conductance. Solid lines present results of Lorentzian fits. (b) Peak conductances $G_\text{max}$ (green) and tunneling rates $\Gamma$ (blue) of the first hole as a function of $V_\mathrm{L,R}$, for dots formed at $V_\mathrm{BG}=\SI{4.8}{V}$ and $V_\mathrm{SG}=\SI{-3.44}{V}$ (circles), and $V_\mathrm{BG}=\SI{3.0}{V}$ and $V_\mathrm{SG}=\SI{-2.45}{V}$ (squares).} 
	\label{figpntun}
\end{figure}
 
As well as electronic QD size discussed in the main text, dot-lead tunnel coupling can also be tuned by barrier gate voltages $V_\mathrm{L,R}$. At $V_\mathrm{BG}=\SI{4.8}{V}$, $V_\mathrm{SG}=\SI{-3.44}{V}$, Coulomb resonance peaks of the first hole at different $V_\mathrm{L,R}$ are shown in Fig.~\ref{figpntun}(a). Lorentzian functions are fitted for each peak. Conductance of each trace is scaled with peak conductance $G_\mathrm{max}$ for clarity. Plunger gate voltage $V_\mathrm{M}$ is converted into energy using lever arms $\alpha_\mathrm1$ extracted for the first hole from Coulomb diamonds measured at corresponding gate voltage configurations. It is then scaled with $k_\mathrm{B}$. At more positive $V_\mathrm{L,R}$ the dot is more strongly coupled to the leads due to the more transparent tunnel barriers as discussed in the main text, giving rise to wider peaks, and vice versa.

From the Lorentzian fits, peak conductance $G_\text{max}$ and full width half maxima (FWHM) (as a measure of tunneling rates) are extracted. The latter is converted into energies (and then into frequencies) using lever arms $\alpha_\mathrm{1}$. The result is plotted in Fig.~\ref{figpntun}(b): with more positive $V_\mathrm{L,R}$, peak conductance $G_\text{max}$ increases from $\sim \SI{}{10^{-3}e^2/h}$ to $\sim e^2/h$ by three orders of magnitudes, and tunneling rates $\Gamma$ from $\sim \SI{10}{GHZ}$ to $\sim \SI{100}{GHz}$, indicating a more strongly coupled dot to the leads.

\subsection{Single dot charge stability diagram}
\label{subsection:VbVp}
\begin{figure}
	\includegraphics[width=8.6cm]{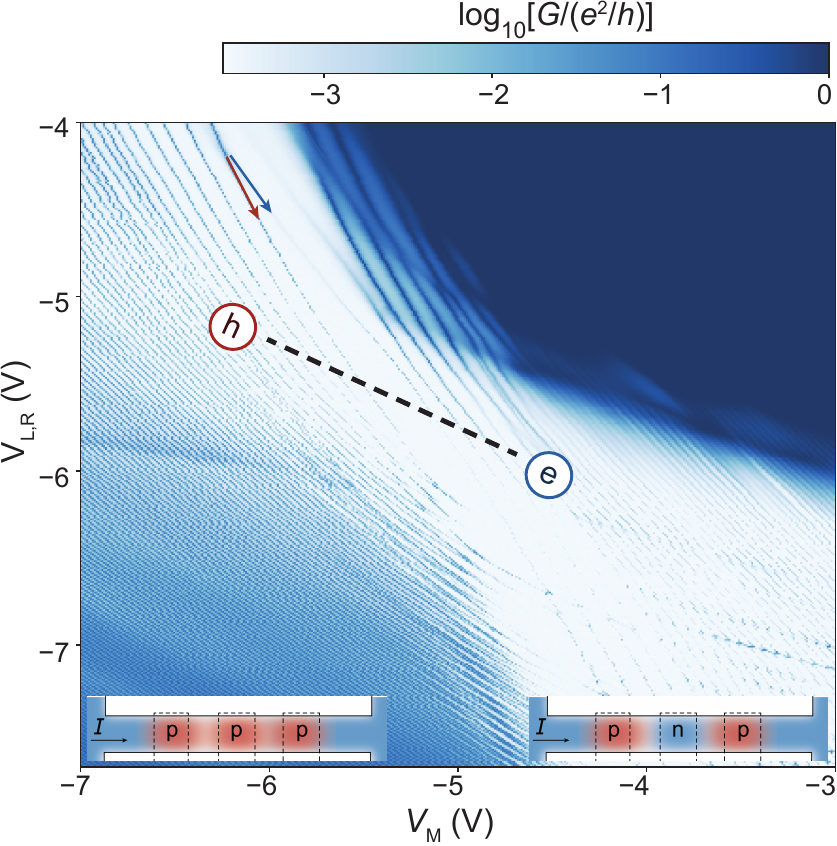}
	\caption{Conductance map of varying plunger gate voltage $V_\mathrm{M}$ and barrier gate voltages $V_\mathrm{L,R}$, at $V_\mathrm{BG}=\SI{3.7}{V}$ and $V_\mathrm{SG}=\SI{-2.92}{V}$. The line along which the Coulomb diamonds in Fig.~\ref{figehsing}(a) is taken is indicated with dashes. Coulomb resonance peaks corresponding to the first hole and TO the first electron states are indicated by the red and the blue arrows, respectively. Insets at the bottom show situations for the corresponding regions, where a large $p$-type dot (bottom left) and triple $p$-$n$-$p$ dots in series (bottom right) is formed. }
	\label{figVbVp}
\end{figure}

Figure.~\ref{figVbVp} shows the conductance map, where barrier gate voltages $V_\mathrm{L,R}$ are swept together with the plunger gate voltage $V_\mathrm{M}$, at $V_\mathrm{BG}=\SI{3.7}{V}$ and $V_\mathrm{SG}=\SI{-2.92}{V}$. At less negative $V_\mathrm{L,R} > \SI{-4}{V}$, the system is open with a $n$-type channel. Barrier gates L and R lower the electron density, but are not enough to deplete regions underneath them into the gap. When $V_\mathrm{L,R}$ become more negative, corresponding to the upper half of the conductance map, the BLG gap $\Delta_\text{gap}$ is opened and $E_\text{F}$ is tuned into this gap, isolating charges from the leads. Applying more negative $V_\mathrm{M}$ from right to left on the map, electrons are removed consecutively from the $n$-type dot until depletion, and consecutive Coulomb resonances can be seen, with the first electron state indicated by the blue arrow. This situation is depicted in Fig.~\ref{figehsing}(b,ii). Starting from depletion, more negative $V_\mathrm{M}$ consecutively adds holes into the now $p$-type dot, depicted in Fig.~\ref{figehsing}(b,i). The red arrow point to the first hole state. When barriers are biased more negatively, in the lower half of the map, they too drive the BLG beneath them into $p$-type, forming a series of $p$-$n$-$p$ triple dots when $V_\mathrm{M}$ is less negative at the bottom right of the map; and forming a large $p$-type dot together with gate M when $V_\mathrm{M}$ is more negative, at the bottom left of the map. The two insets depict the situations at corresponding locations of the map. 

Instead of being completely perpendicular to the $V_\mathrm{M}$-axis or to the $V_\mathrm{L,R}$-axis, the Coulomb resonance lines come at an angle, as can be seen from the map, demonstrating the finite cross-talk between the barrier gates L and R and the plunger gate M. This cross-talk stems from the close separation (75nm) between gates L, M, and R. To mitigate the effect of such cross-talk and  to keep the barrier strength roughly constant, we perform bipolar operation [shown in Fig.~\ref{figehsing}(a)] at the dashed line on the map.

\section{Theoretical models}
\label{section:Theory}
\renewcommand\theequation{\Alph{section}\arabic{equation}} 
\begin{figure*}
	\includegraphics[width=\textwidth]{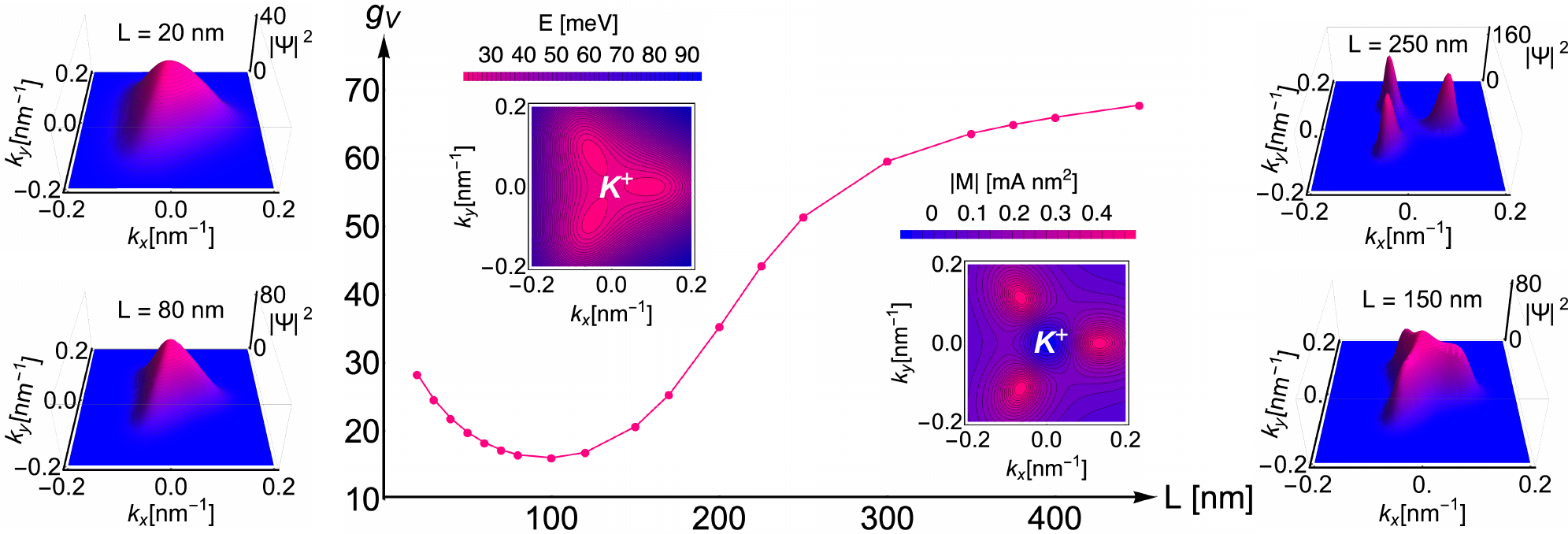}
	\caption{The dot wave functions in the $K^{+}$ valley and valley g-factor as a function of the dot size for a circularly symmetric dot with a gap of $\Delta=\SI{42}{meV}$ at the center. The insets show the corresponding BLG dispersion, E, of the first conduction band (top) and orbital magnetic moment, M (bottom) in valley $K^+$.}
	\label{fig4}
\end{figure*}

We model the BLG QD, formed with the help of electrostatic  split gates \cite{mariusnanolett, Hiske2018electrostatically, Overweg2018a, angelika2020quartet}, by a smooth confinement potential, $U(\mathbf{r})$, and a gap profile, $\Delta(\mathbf{r})$, which enter in the single-electron four-band Hamiltonian \cite{McCann2007, McCann2013},
\begin{align}
 \nonumber H_{\pm}= &
\begin{pmatrix} 
 U \mp\frac{1}{2}\Delta  & \pm v_3\pi & 0 &\pm v \pi^{\dagger}\\
\pm v_3 \pi^{\dagger}&  U \pm\frac{1}{2}\Delta  & \pm v\pi &0\\
 0 & \pm v\pi^{\dagger} &   U \pm\frac{1}{2}\Delta  &   \gamma_1\\
\pm v\pi & 0 &   \gamma_1 &  U \mp\frac{1}{2}\Delta 
\end{pmatrix},\\
\nonumber & U(x,y)= \frac{U_0}{\cosh{\frac{\sqrt{(\frac{x}{a})^2+(\frac{y}{b})^2}}{L}}}, \\
 &\Delta(x,y)= \Delta_0-\frac{0.3\Delta_0}{\cosh{\frac{\sqrt{(\frac{x}{a})^2+(\frac{y}{b})^2}}{L}}}.
\label{eqn:H}
\end{align}
with momenta $\pi=p_x+ip_y,\,  \pi^{\dagger}=p_x-ip_y$ and velocities $v=1.02*10^6 \text{ m/s}$ and  $v_3\approx0.12 v$, and energy $\gamma_1\approx0.38\text{ eV}$. This Hamiltonian  is written for the Bloch function components $\psi_{K^+}=(\psi_{A},\psi_{B^{\prime}},\psi_{A^{\prime}},\psi_{B})$ in valley $K^+$, and $\psi_{K^-}=(\psi_{B^{\prime}},\psi_{A},\psi_{B},\psi_{A^{\prime}})$ in valley $K^-$,  with electron's amplitudes on the BLG sublattices $A$ and $B$ in the top, and  $A^{\prime}$ and $B^{\prime}$ in the bottom layer.

In the absence of confinement, Eq.~\ref{eqn:H} describes the low energy trigonally warped bands \cite{Varlet2014, Varlet2015, angelika2018minivalley} featuring three mini-valleys around each $K$ point (Fig.~\ref{fig4} top inset). The Bloch bands of gapped BLG have a non-trivial Berry curvature which gives rise to an orbital magnetic moment,  $\mathbf{M}(\mathbf{k})=M(\mathbf{k})\mathbf{e}_z$, \cite{Xiao2010, Chang1996a}
\begin{equation}
{M}=-i\frac{e}{2\hbar}\langle\mathbf{\nabla}_{\mathbf{k}}\Phi(\mathbf{k})|\times [\epsilon(\mathbf{k})-H(\mathbf{k})] |\mathbf{\nabla}_{\mathbf{k}}\Phi(\mathbf{k})\rangle\cdot\mathbf{e}_z,
\label{eqn:Berry}
\end{equation}
where, $\mathbf{\nabla}_{\mathbf{k}}=(\partial_{k_x},\partial_{k_y})$, $"\times"$ is the cross product,  $\epsilon(\mathbf{k})$ is the band energy, and $\Phi$ is the corresponding Bloch state. The orbital magnetic moment is maximal around the band edges, i.e., it is peaked around the mini-valleys of the trigonally warped gapped BLG dispersion and carries opposite sign in the two different valleys (bottom inset in Fig.~\ref{fig4}). The coupling of the orbital magnetic moment to a magnetic field dominates the valley splitting of the dot states in weak magnetic fields \cite{angelika2020quartet, Lee20}.

To obtain the electronic structure of the confined BLG QD we diagonalize the Hamiltonian,  $H_{\pm}$, in Eq.~\ref{eqn:H} numerically in a suitable basis of localized states. We choose the eigenstates of the two-dimensional  harmonic oscillator  (products of wave functions $\psi_n(x)=N_n e^{-\frac{1}{2}(\alpha x)^2}\mathcal{H}_n(\alpha x)$, where $N_n= \sqrt{\frac{\alpha}{\sqrt{\pi}2^n n!}}$ is the normalization constant and $\alpha$ is a scaling factor of unit length$^{-1}$; we adapt $\alpha$ to the potential $U(x)$ mimicked by a parabolic potential at the bottom of $U$). The basis states are then given by 
\begin{equation}\psi_{\eta\mu ,1}=
\nonumber\begin{pmatrix}
\psi_{\eta}(x)\psi_{\mu}(y)\\
0\\
0\\
0
\end{pmatrix},\;
\psi_{\eta\mu,2}=
\begin{pmatrix}
0\\
\psi_{\eta}(x)\psi_{\mu}(y)\\
0\\
0
\end{pmatrix},
\end{equation}
\begin{equation}
\psi_{\eta\mu,3}-
\begin{pmatrix}
0\\
0\\
\psi_{\eta}(x)\psi_{\mu}(y)\\
0
\end{pmatrix},
\psi_{ \eta\mu,4}=
\begin{pmatrix}
0\\
0\\
0\\
\psi_{\eta}(x)\psi_{\mu}(y).
\end{pmatrix}.
 \label{eqn:BasisHarm}
\end{equation}
For every set of system parameters we construct the matrix corresponding to Hamiltonian $H_{\pm}$ in the basis given in Eq.~\ref{eqn:BasisHarm} and obtain the energy spectrum by diagonalization. The spectrum is converged when the energy levels change no more upon including a higher number of basis states.

For the lowest single-particle dot state, $\Psi$, we compute its valley g-factor, $g_\mathrm{v}$, by estimating how much orbital angular momentum is picked up in momentum space \cite{Lee20, angelika2020quartet}
\begin{equation}
{g}_v=\frac{2}{\mu_B}\int d\boldsymbol{k} M_z(\boldsymbol{k} )|\Psi(\boldsymbol{k} )|^2,
\label{eqn:gV}
\end{equation}
where $\mu_B$ is the Bohr magneton. The factor 2 in Eq.~\ref{eqn:gV} stems from the fact that the total valley splitting $\Delta E_\mathrm{K^-,K^+}$ is twice that of a single valley as $M$ is of equal magnitude but opposite sign in the two valleys. 

For a circularly symmetric dot ($a\equiv b$ in Eq.~\ref{eqn:H}), we show the dependence of $g_\mathrm{v}$ on the dot size in Fig.~\ref{fig4} alongside with the momentum space distribution of the lowest dot state, $\Psi$. In this figure, we chose the parameters $U_0=\SI{-2}{meV}$ and $\Delta_0=\SI{60}{meV}$ (which amounts to a gap of $\Delta = \SI{42}{meV}$ at the dot's center) in Eq.~\ref{eqn:H}, as well as $a=b=1$, while varying $L$. For small dots, increasing the dot size reduces the support of the wave function in momentum space. Consequently, the states pick up less orbital momentum in Eq.~\ref{eqn:gV} and the valley g-factor reduces. For larger dots, the bottom of the band is more shallow and the three mini-valleys of the BLG dispersion influence the dot state. In this regime, the larger the dot, the more the state is squeezed into the mini-valleys, where the orbital magnetic moment is maximal. Hence the valley g-factor increases with dot size. The latter regime is favored by large dot sizes, shallow confinement potentials and large gaps in the BLG dispersion \cite{angelika2020quartet}. 

To model elliptically elongated dots in the $x$ or $y$ direction, we vary the parameters $a$ and $b$ in Eq.~\ref{eqn:H}, respectively. In an elliptical QD, the ground state's wave functions are squeezed mainly into one (elongation along the $x$-axis) or two (elongation along the $y$-axis) of the three mini-valleys, leading to a further increase of the valley g-factor.


\begin{thebibliography}{33}%
\makeatletter
\providecommand \@ifxundefined [1]{%
 \@ifx{#1\undefined}
}%
\providecommand \@ifnum [1]{%
 \ifnum #1\expandafter \@firstoftwo
 \else \expandafter \@secondoftwo
 \fi
}%
\providecommand \@ifx [1]{%
 \ifx #1\expandafter \@firstoftwo
 \else \expandafter \@secondoftwo
 \fi
}%
\providecommand \natexlab [1]{#1}%
\providecommand \enquote  [1]{``#1''}%
\providecommand \bibnamefont  [1]{#1}%
\providecommand \bibfnamefont [1]{#1}%
\providecommand \citenamefont [1]{#1}%
\providecommand \href@noop [0]{\@secondoftwo}%
\providecommand \href [0]{\begingroup \@sanitize@url \@href}%
\providecommand \@href[1]{\@@startlink{#1}\@@href}%
\providecommand \@@href[1]{\endgroup#1\@@endlink}%
\providecommand \@sanitize@url [0]{\catcode `\\12\catcode `\$12\catcode
  `\&12\catcode `\#12\catcode `\^12\catcode `\_12\catcode `\%12\relax}%
\providecommand \@@startlink[1]{}%
\providecommand \@@endlink[0]{}%
\providecommand \url  [0]{\begingroup\@sanitize@url \@url }%
\providecommand \@url [1]{\endgroup\@href {#1}{\urlprefix }}%
\providecommand \urlprefix  [0]{URL }%
\providecommand \Eprint [0]{\href }%
\providecommand \doibase [0]{https://doi.org/}%
\providecommand \selectlanguage [0]{\@gobble}%
\providecommand \bibinfo  [0]{\@secondoftwo}%
\providecommand \bibfield  [0]{\@secondoftwo}%
\providecommand \translation [1]{[#1]}%
\providecommand \BibitemOpen [0]{}%
\providecommand \bibitemStop [0]{}%
\providecommand \bibitemNoStop [0]{.\EOS\space}%
\providecommand \EOS [0]{\spacefactor3000\relax}%
\providecommand \BibitemShut  [1]{\csname bibitem#1\endcsname}%
\let\auto@bib@innerbib\@empty
\bibitem [{\citenamefont {Sze}(1969)}]{Sze69}%
  \BibitemOpen
  \bibfield  {author} {\bibinfo {author} {\bibfnamefont {S.~M.}\ \bibnamefont
  {Sze}},\ }\href@noop {} {\emph {\bibinfo {title} {Physics of Semiconductor
  Devices}}}\ (\bibinfo  {publisher} {Wiley},\ \bibinfo {address} {New York},\
  \bibinfo {year} {1969})\BibitemShut {NoStop}%
\bibitem [{\citenamefont {Shayegan}\ \emph {et~al.}(2006)\citenamefont
  {Shayegan}, \citenamefont {De~Poortere}, \citenamefont {Gunawan},
  \citenamefont {Shkolnikov}, \citenamefont {Tutuc},\ and\ \citenamefont
  {Vakili}}]{Shayegan06}%
  \BibitemOpen
  \bibfield  {author} {\bibinfo {author} {\bibfnamefont {M.}~\bibnamefont
  {Shayegan}}, \bibinfo {author} {\bibfnamefont {E.~P.}\ \bibnamefont
  {De~Poortere}}, \bibinfo {author} {\bibfnamefont {O.}~\bibnamefont
  {Gunawan}}, \bibinfo {author} {\bibfnamefont {Y.~P.}\ \bibnamefont
  {Shkolnikov}}, \bibinfo {author} {\bibfnamefont {E.}~\bibnamefont {Tutuc}},\
  and\ \bibinfo {author} {\bibfnamefont {K.}~\bibnamefont {Vakili}},\
  }\bibfield  {title} {\bibinfo {title} {Two-dimensional electrons occupying
  multiple valleys in alas},\ }\href {https://doi.org/10.1002/pssb.200642212}
  {\bibfield  {journal} {\bibinfo  {journal} {Phys. Stat. Sol.}\ }\textbf
  {\bibinfo {volume} {(b) 243}},\ \bibinfo {pages} {3629 – 3642} (\bibinfo
  {year} {2006})}\BibitemShut {NoStop}%
\bibitem [{\citenamefont {Schaibley}\ \emph {et~al.}(2016)\citenamefont
  {Schaibley}, \citenamefont {Yu}, \citenamefont {Clark}, \citenamefont
  {Rivera}, \citenamefont {Ross}, \citenamefont {Seyler}, \citenamefont {Yao},\
  and\ \citenamefont {Xu}}]{Schaibley16}%
  \BibitemOpen
  \bibfield  {author} {\bibinfo {author} {\bibfnamefont {J.~R.}\ \bibnamefont
  {Schaibley}}, \bibinfo {author} {\bibfnamefont {H.}~\bibnamefont {Yu}},
  \bibinfo {author} {\bibfnamefont {G.}~\bibnamefont {Clark}}, \bibinfo
  {author} {\bibfnamefont {P.}~\bibnamefont {Rivera}}, \bibinfo {author}
  {\bibfnamefont {J.~S.}\ \bibnamefont {Ross}}, \bibinfo {author}
  {\bibfnamefont {K.~L.}\ \bibnamefont {Seyler}}, \bibinfo {author}
  {\bibfnamefont {W.}~\bibnamefont {Yao}},\ and\ \bibinfo {author}
  {\bibfnamefont {X.}~\bibnamefont {Xu}},\ }\bibfield  {title} {\bibinfo
  {title} {Valleytronics in 2d materials},\ }\bibfield  {journal} {\bibinfo
  {journal} {Nature Reviews Materials}\ }\textbf {\bibinfo {volume} {1}},\
  \href {https://doi.org/10.1038/natrevmats.2016.55}
  {10.1038/natrevmats.2016.55} (\bibinfo {year} {2016})\BibitemShut {NoStop}%
\bibitem [{\citenamefont {Thompson}\ \emph {et~al.}(2004)\citenamefont
  {Thompson}, \citenamefont {Armstrong}, \citenamefont {Auth}, \citenamefont
  {Alavi}, \citenamefont {Buehler}, \citenamefont {Chau}, \citenamefont {Cea},
  \citenamefont {Ghani}, \citenamefont {Glass}, \citenamefont {Hoffman} \emph
  {et~al.}}]{Thompson04}%
  \BibitemOpen
  \bibfield  {author} {\bibinfo {author} {\bibfnamefont {S.~E.}\ \bibnamefont
  {Thompson}}, \bibinfo {author} {\bibfnamefont {M.}~\bibnamefont {Armstrong}},
  \bibinfo {author} {\bibfnamefont {C.}~\bibnamefont {Auth}}, \bibinfo {author}
  {\bibfnamefont {M.}~\bibnamefont {Alavi}}, \bibinfo {author} {\bibfnamefont
  {M.}~\bibnamefont {Buehler}}, \bibinfo {author} {\bibfnamefont
  {R.}~\bibnamefont {Chau}}, \bibinfo {author} {\bibfnamefont {S.}~\bibnamefont
  {Cea}}, \bibinfo {author} {\bibfnamefont {T.}~\bibnamefont {Ghani}}, \bibinfo
  {author} {\bibfnamefont {G.}~\bibnamefont {Glass}}, \bibinfo {author}
  {\bibfnamefont {T.}~\bibnamefont {Hoffman}}, \emph {et~al.},\ }\bibfield
  {title} {\bibinfo {title} {A 90-nm logic technology featuring
  strained-silicon},\ }\href {https://doi.org/10.1109/TED.2004.836648}
  {\bibfield  {journal} {\bibinfo  {journal} {IEEE Transactions on electron
  devices}\ }\textbf {\bibinfo {volume} {51}},\ \bibinfo {pages} {1790}
  (\bibinfo {year} {2004})}\BibitemShut {NoStop}%
\bibitem [{\citenamefont {Rohling}\ and\ \citenamefont
  {Burkard}(2012)}]{Rohling12}%
  \BibitemOpen
  \bibfield  {author} {\bibinfo {author} {\bibfnamefont {N.}~\bibnamefont
  {Rohling}}\ and\ \bibinfo {author} {\bibfnamefont {G.}~\bibnamefont
  {Burkard}},\ }\bibfield  {title} {\bibinfo {title} {Universal quantum
  computing with spin and valley states},\ }\href
  {https://doi.org/10.1088/1367-2630/14/8/083008} {\bibfield  {journal}
  {\bibinfo  {journal} {New J. Phys.}\ }\textbf {\bibinfo {volume} {14}},\
  \bibinfo {pages} {083008} (\bibinfo {year} {2012})}\BibitemShut {NoStop}%
\bibitem [{\citenamefont {Penthorn}\ \emph {et~al.}(2019)\citenamefont
  {Penthorn}, \citenamefont {Schoenfield}, \citenamefont {Rooney},
  \citenamefont {Edge},\ and\ \citenamefont {Jiang}}]{Penthorn19}%
  \BibitemOpen
  \bibfield  {author} {\bibinfo {author} {\bibfnamefont {N.~E.}\ \bibnamefont
  {Penthorn}}, \bibinfo {author} {\bibfnamefont {J.~S.}\ \bibnamefont
  {Schoenfield}}, \bibinfo {author} {\bibfnamefont {J.~D.}\ \bibnamefont
  {Rooney}}, \bibinfo {author} {\bibfnamefont {L.~F.}\ \bibnamefont {Edge}},\
  and\ \bibinfo {author} {\bibfnamefont {H.}~\bibnamefont {Jiang}},\ }\bibfield
   {title} {\bibinfo {title} {Two-axis quantum control of a fast valley qubit
  in silicon},\ }\href {https://doi.org/10.1038/s41534-019-0212-5} {\bibfield
  {journal} {\bibinfo  {journal} {npj Quantum Information}\ }\textbf {\bibinfo
  {volume} {5}},\ \bibinfo {pages} {1} (\bibinfo {year} {2019})}\BibitemShut
  {NoStop}%
\bibitem [{\citenamefont {Loss}\ and\ \citenamefont
  {DiVincenzo}(1998)}]{Loss98}%
  \BibitemOpen
  \bibfield  {author} {\bibinfo {author} {\bibfnamefont {D.}~\bibnamefont
  {Loss}}\ and\ \bibinfo {author} {\bibfnamefont {D.~P.}\ \bibnamefont
  {DiVincenzo}},\ }\bibfield  {title} {\bibinfo {title} {Quantum computation
  with quantum dots},\ }\href {https://doi.org/10.1103/PhysRevA.57.120}
  {\bibfield  {journal} {\bibinfo  {journal} {Phys. Rev. A}\ }\textbf {\bibinfo
  {volume} {57}},\ \bibinfo {pages} {120} (\bibinfo {year} {1998})}\BibitemShut
  {NoStop}%
\bibitem [{\citenamefont {Salis}\ \emph {et~al.}(2001)\citenamefont {Salis},
  \citenamefont {Kato}, \citenamefont {Ensslin}, \citenamefont {Driscoll},
  \citenamefont {Gossard},\ and\ \citenamefont {Awschalom}}]{Salis01}%
  \BibitemOpen
  \bibfield  {author} {\bibinfo {author} {\bibfnamefont {G.}~\bibnamefont
  {Salis}}, \bibinfo {author} {\bibfnamefont {Y.}~\bibnamefont {Kato}},
  \bibinfo {author} {\bibfnamefont {K.}~\bibnamefont {Ensslin}}, \bibinfo
  {author} {\bibfnamefont {D.}~\bibnamefont {Driscoll}}, \bibinfo {author}
  {\bibfnamefont {A.}~\bibnamefont {Gossard}},\ and\ \bibinfo {author}
  {\bibfnamefont {D.}~\bibnamefont {Awschalom}},\ }\bibfield  {title} {\bibinfo
  {title} {Electrical control of spin coherence in semiconductor
  nanostructures},\ }\href {https://doi.org/10.1038/414619a} {\bibfield
  {journal} {\bibinfo  {journal} {Nature}\ }\textbf {\bibinfo {volume} {414}},\
  \bibinfo {pages} {619} (\bibinfo {year} {2001})}\BibitemShut {NoStop}%
\bibitem [{\citenamefont {Kato}\ \emph {et~al.}(2003)\citenamefont {Kato},
  \citenamefont {Myers}, \citenamefont {Driscoll}, \citenamefont {Gossard},
  \citenamefont {Levy},\ and\ \citenamefont {Awschalom}}]{Kato03}%
  \BibitemOpen
  \bibfield  {author} {\bibinfo {author} {\bibfnamefont {Y.}~\bibnamefont
  {Kato}}, \bibinfo {author} {\bibfnamefont {R.}~\bibnamefont {Myers}},
  \bibinfo {author} {\bibfnamefont {D.}~\bibnamefont {Driscoll}}, \bibinfo
  {author} {\bibfnamefont {A.}~\bibnamefont {Gossard}}, \bibinfo {author}
  {\bibfnamefont {J.}~\bibnamefont {Levy}},\ and\ \bibinfo {author}
  {\bibfnamefont {D.}~\bibnamefont {Awschalom}},\ }\bibfield  {title} {\bibinfo
  {title} {Gigahertz electron spin manipulation using voltage-controlled
  g-tensor modulation},\ }\href {https://doi.org/10.1126/science.1080880}
  {\bibfield  {journal} {\bibinfo  {journal} {Science}\ }\textbf {\bibinfo
  {volume} {299}},\ \bibinfo {pages} {1201} (\bibinfo {year}
  {2003})}\BibitemShut {NoStop}%
\bibitem [{\citenamefont {Nitta}\ \emph {et~al.}(2003)\citenamefont {Nitta},
  \citenamefont {Lin}, \citenamefont {Akazaki},\ and\ \citenamefont
  {Koga}}]{Nitta03}%
  \BibitemOpen
  \bibfield  {author} {\bibinfo {author} {\bibfnamefont {J.}~\bibnamefont
  {Nitta}}, \bibinfo {author} {\bibfnamefont {Y.}~\bibnamefont {Lin}}, \bibinfo
  {author} {\bibfnamefont {T.}~\bibnamefont {Akazaki}},\ and\ \bibinfo {author}
  {\bibfnamefont {T.}~\bibnamefont {Koga}},\ }\bibfield  {title} {\bibinfo
  {title} {Gate-controlled electron g factor in an inas-inserted-channel in
  0.53 ga 0.47 as/in 0.52 al 0.48 as heterostructure},\ }\href
  {https://doi.org/10.1063/1.1631082} {\bibfield  {journal} {\bibinfo
  {journal} {Appl. Phys. Lett.}\ }\textbf {\bibinfo {volume} {83}},\ \bibinfo
  {pages} {4565} (\bibinfo {year} {2003})}\BibitemShut {NoStop}%
\bibitem [{\citenamefont {Bj{\"o}rk}\ \emph {et~al.}(2005)\citenamefont
  {Bj{\"o}rk}, \citenamefont {Fuhrer}, \citenamefont {Hansen}, \citenamefont
  {Larsson}, \citenamefont {Fr{\"o}berg},\ and\ \citenamefont
  {Samuelson}}]{Bjork05}%
  \BibitemOpen
  \bibfield  {author} {\bibinfo {author} {\bibfnamefont {M.~T.}\ \bibnamefont
  {Bj{\"o}rk}}, \bibinfo {author} {\bibfnamefont {A.}~\bibnamefont {Fuhrer}},
  \bibinfo {author} {\bibfnamefont {A.~E.}\ \bibnamefont {Hansen}}, \bibinfo
  {author} {\bibfnamefont {M.~W.}\ \bibnamefont {Larsson}}, \bibinfo {author}
  {\bibfnamefont {L.~E.}\ \bibnamefont {Fr{\"o}berg}},\ and\ \bibinfo {author}
  {\bibfnamefont {L.}~\bibnamefont {Samuelson}},\ }\bibfield  {title} {\bibinfo
  {title} {Tunable effective g factor in inas nanowire quantum dots},\ }\href
  {https://doi.org/10.1103/PhysRevB.72.201307} {\bibfield  {journal} {\bibinfo
  {journal} {Phys. Rev. B}\ }\textbf {\bibinfo {volume} {72}},\ \bibinfo
  {pages} {201307(R)} (\bibinfo {year} {2005})}\BibitemShut {NoStop}%
\bibitem [{\citenamefont {Hollmann}\ \emph {et~al.}(2020)\citenamefont
  {Hollmann}, \citenamefont {Struck}, \citenamefont {Langrock}, \citenamefont
  {Schmidbauer}, \citenamefont {Schauer}, \citenamefont {Leonhardt},
  \citenamefont {Sawano}, \citenamefont {Riemann}, \citenamefont {Abrosimov},
  \citenamefont {Bougeard} \emph {et~al.}}]{hollmann2020large}%
  \BibitemOpen
  \bibfield  {author} {\bibinfo {author} {\bibfnamefont {A.}~\bibnamefont
  {Hollmann}}, \bibinfo {author} {\bibfnamefont {T.}~\bibnamefont {Struck}},
  \bibinfo {author} {\bibfnamefont {V.}~\bibnamefont {Langrock}}, \bibinfo
  {author} {\bibfnamefont {A.}~\bibnamefont {Schmidbauer}}, \bibinfo {author}
  {\bibfnamefont {F.}~\bibnamefont {Schauer}}, \bibinfo {author} {\bibfnamefont
  {T.}~\bibnamefont {Leonhardt}}, \bibinfo {author} {\bibfnamefont
  {K.}~\bibnamefont {Sawano}}, \bibinfo {author} {\bibfnamefont
  {H.}~\bibnamefont {Riemann}}, \bibinfo {author} {\bibfnamefont {N.~V.}\
  \bibnamefont {Abrosimov}}, \bibinfo {author} {\bibfnamefont {D.}~\bibnamefont
  {Bougeard}}, \emph {et~al.},\ }\bibfield  {title} {\bibinfo {title} {Large,
  tunable valley splitting and single-spin relaxation mechanisms in a si/six
  ge1-x quantum dot},\ }\href
  {https://doi.org/10.1103/PhysRevApplied.13.034068} {\bibfield  {journal}
  {\bibinfo  {journal} {Physical Review Applied}\ }\textbf {\bibinfo {volume}
  {13}},\ \bibinfo {pages} {034068} (\bibinfo {year} {2020})}\BibitemShut
  {NoStop}%
\bibitem [{\citenamefont {Trauzettel}\ \emph {et~al.}(2007)\citenamefont
  {Trauzettel}, \citenamefont {Bulaev}, \citenamefont {Loss},\ and\
  \citenamefont {Burkard}}]{Trauzettel2007spinqubit}%
  \BibitemOpen
  \bibfield  {author} {\bibinfo {author} {\bibfnamefont {B.}~\bibnamefont
  {Trauzettel}}, \bibinfo {author} {\bibfnamefont {D.~V.}\ \bibnamefont
  {Bulaev}}, \bibinfo {author} {\bibfnamefont {D.}~\bibnamefont {Loss}},\ and\
  \bibinfo {author} {\bibfnamefont {G.}~\bibnamefont {Burkard}},\ }\bibfield
  {title} {\bibinfo {title} {Spin qubits in graphene quantum dots},\ }\href
  {https://doi.org/10.1038/nphys544} {\bibfield  {journal} {\bibinfo  {journal}
  {Nature Phys.}\ }\textbf {\bibinfo {volume} {3}},\ \bibinfo {pages} {192}
  (\bibinfo {year} {2007})}\BibitemShut {NoStop}%
\bibitem [{\citenamefont {Lee}\ \emph {et~al.}(2020)\citenamefont {Lee},
  \citenamefont {Knothe}, \citenamefont {Overweg}, \citenamefont {Eich},
  \citenamefont {Gold}, \citenamefont {Kurzmann}, \citenamefont {Klasovika},
  \citenamefont {Taniguchi}, \citenamefont {Wantanabe}, \citenamefont
  {Fal’ko} \emph {et~al.}}]{Lee20}%
  \BibitemOpen
  \bibfield  {author} {\bibinfo {author} {\bibfnamefont {Y.}~\bibnamefont
  {Lee}}, \bibinfo {author} {\bibfnamefont {A.}~\bibnamefont {Knothe}},
  \bibinfo {author} {\bibfnamefont {H.}~\bibnamefont {Overweg}}, \bibinfo
  {author} {\bibfnamefont {M.}~\bibnamefont {Eich}}, \bibinfo {author}
  {\bibfnamefont {C.}~\bibnamefont {Gold}}, \bibinfo {author} {\bibfnamefont
  {A.}~\bibnamefont {Kurzmann}}, \bibinfo {author} {\bibfnamefont
  {V.}~\bibnamefont {Klasovika}}, \bibinfo {author} {\bibfnamefont
  {T.}~\bibnamefont {Taniguchi}}, \bibinfo {author} {\bibfnamefont
  {K.}~\bibnamefont {Wantanabe}}, \bibinfo {author} {\bibfnamefont
  {V.}~\bibnamefont {Fal’ko}}, \emph {et~al.},\ }\bibfield  {title} {\bibinfo
  {title} {Tunable valley splitting due to topological orbital magnetic moment
  in bilayer graphene quantum point contacts},\ }\href
  {https://doi.org/10.1103/PhysRevLett.124.126802} {\bibfield  {journal}
  {\bibinfo  {journal} {Phys. Rev. Lett.}\ }\textbf {\bibinfo {volume} {124}},\
  \bibinfo {pages} {126802} (\bibinfo {year} {2020})}\BibitemShut {NoStop}%
\bibitem [{\citenamefont {Knothe}\ and\ \citenamefont
  {Fal'ko}(2018)}]{angelika2018minivalley}%
  \BibitemOpen
  \bibfield  {author} {\bibinfo {author} {\bibfnamefont {A.}~\bibnamefont
  {Knothe}}\ and\ \bibinfo {author} {\bibfnamefont {V.}~\bibnamefont
  {Fal'ko}},\ }\bibfield  {title} {\bibinfo {title} {Influence of minivalleys
  and berry curvature on electrostatically induced quantum wires in gapped
  bilayer graphene},\ }\href {https://doi.org/10.1103/PhysRevB.98.155435}
  {\bibfield  {journal} {\bibinfo  {journal} {Phys. Rev. B}\ }\textbf {\bibinfo
  {volume} {98}},\ \bibinfo {pages} {155435} (\bibinfo {year}
  {2018})}\BibitemShut {NoStop}%
\bibitem [{\citenamefont {Knothe}\ and\ \citenamefont
  {Fal'ko}(2020)}]{angelika2020quartet}%
  \BibitemOpen
  \bibfield  {author} {\bibinfo {author} {\bibfnamefont {A.}~\bibnamefont
  {Knothe}}\ and\ \bibinfo {author} {\bibfnamefont {V.}~\bibnamefont
  {Fal'ko}},\ }\bibfield  {title} {\bibinfo {title} {Quartet states in
  two-electron quantum dots in bilayer graphene},\ }\href
  {https://doi.org/10.1103/PhysRevB.101.235423} {\bibfield  {journal} {\bibinfo
   {journal} {Phys. Rev. B}\ }\textbf {\bibinfo {volume} {101}},\ \bibinfo
  {pages} {235423} (\bibinfo {year} {2020})}\BibitemShut {NoStop}%
\bibitem [{\citenamefont {Eich}\ \emph
  {et~al.}(2018{\natexlab{a}})\citenamefont {Eich}, \citenamefont {Herman},\citenamefont {Pisoni},
  \citenamefont {Overweg}, \citenamefont {Kurzmann}, \citenamefont {Lee},
  \citenamefont {Rickhaus}, \citenamefont {Ihn}, \citenamefont {Ensslin}, \citenamefont {Sigrist} \emph {et~al.}}]{Mariusprx}%
  \BibitemOpen
  \bibfield  {author} {\bibinfo {author} {\bibfnamefont {M.}~\bibnamefont
  {Eich}}, \bibinfo {author} {\bibfnamefont {F.}~\bibnamefont {Herman}}, \bibinfo {author} {\bibfnamefont {R.}~\bibnamefont {Pisoni}},
  \bibinfo {author} {\bibfnamefont {H.}~\bibnamefont {Overweg}}, \bibinfo
  {author} {\bibfnamefont {A.}~\bibnamefont {Kurzmann}}, \bibinfo {author}
  {\bibfnamefont {Y.}~\bibnamefont {Lee}}, \bibinfo {author} {\bibfnamefont
  {P.}~\bibnamefont {Rickhaus}}, \bibinfo {author} {\bibfnamefont
  {T.}~\bibnamefont {Ihn}}, \bibinfo {author} {\bibfnamefont {K.}~\bibnamefont
  {Ensslin}}, \bibinfo {author} {\bibfnamefont {M.}~\bibnamefont {Sigrist}}, \emph
  {et~al.},\ }\bibfield  {title} {\bibinfo {title} {Spin and valley states in
  gate-defined bilayer graphene quantum dots},\ }\href
  {https://doi.org/10.1103/PhysRevX.8.031023} {\bibfield  {journal} {\bibinfo
  {journal} {Phys. Rev. X}\ }\textbf {\bibinfo {volume} {8}},\ \bibinfo {pages}
  {031023} (\bibinfo {year} {2018}{\natexlab{a}})}\BibitemShut {NoStop}%
\bibitem [{\citenamefont {Kurzmann}\ \emph {et~al.}(2019)\citenamefont
  {Kurzmann}, \citenamefont {Eich}, \citenamefont {Overweg}, \citenamefont
  {Mangold}, \citenamefont {Herman}, \citenamefont {Rickhaus}, \citenamefont
  {Pisoni}, \citenamefont {Lee}, \citenamefont {Garreis}, \citenamefont {Tong}
  \emph {et~al.}}]{annikaexcitedstates}%
  \BibitemOpen
  \bibfield  {author} {\bibinfo {author} {\bibfnamefont {A.}~\bibnamefont
  {Kurzmann}}, \bibinfo {author} {\bibfnamefont {M.}~\bibnamefont {Eich}},
  \bibinfo {author} {\bibfnamefont {H.}~\bibnamefont {Overweg}}, \bibinfo
  {author} {\bibfnamefont {M.}~\bibnamefont {Mangold}}, \bibinfo {author}
  {\bibfnamefont {F.}~\bibnamefont {Herman}}, \bibinfo {author} {\bibfnamefont
  {P.}~\bibnamefont {Rickhaus}}, \bibinfo {author} {\bibfnamefont
  {R.}~\bibnamefont {Pisoni}}, \bibinfo {author} {\bibfnamefont
  {Y.}~\bibnamefont {Lee}}, \bibinfo {author} {\bibfnamefont {R.}~\bibnamefont
  {Garreis}}, \bibinfo {author} {\bibfnamefont {C.}~\bibnamefont {Tong}}, \emph
  {et~al.},\ }\bibfield  {title} {\bibinfo {title} {Excited states in bilayer
  graphene quantum dots},\ }\href
  {https://doi.org/10.1103/PhysRevLett.123.026803} {\bibfield  {journal}
  {\bibinfo  {journal} {Phys. Rev. Lett.}\ }\textbf {\bibinfo {volume} {123}},\
  \bibinfo {pages} {026803} (\bibinfo {year} {2019})}\BibitemShut {NoStop}%
\bibitem [{\citenamefont {Ohta}\ \emph {et~al.}(2006)\citenamefont {Ohta},
  \citenamefont {Bostwick}, \citenamefont {Seyller}, \citenamefont {Horn},\
  and\ \citenamefont {Rotenberg}}]{Ohta2006BLGband}%
  \BibitemOpen
  \bibfield  {author} {\bibinfo {author} {\bibfnamefont {T.}~\bibnamefont
  {Ohta}}, \bibinfo {author} {\bibfnamefont {A.}~\bibnamefont {Bostwick}},
  \bibinfo {author} {\bibfnamefont {T.}~\bibnamefont {Seyller}}, \bibinfo
  {author} {\bibfnamefont {K.}~\bibnamefont {Horn}},\ and\ \bibinfo {author}
  {\bibfnamefont {E.}~\bibnamefont {Rotenberg}},\ }\bibfield  {title} {\bibinfo
  {title} {Controlling the electronic structure of bilayer graphene},\ }\href
  {https://doi.org/10.1126/science.1130681} {\bibfield  {journal} {\bibinfo
  {journal} {Science}\ }\textbf {\bibinfo {volume} {313}},\ \bibinfo {pages}
  {951} (\bibinfo {year} {2006})}\BibitemShut {NoStop}%
\bibitem [{\citenamefont {McCann}(2006)}]{mccann2006BLGband}%
  \BibitemOpen
  \bibfield  {author} {\bibinfo {author} {\bibfnamefont {E.}~\bibnamefont
  {McCann}},\ }\bibfield  {title} {\bibinfo {title} {Asymmetry gap in the
  electronic band structure of bilayer graphene},\ }\href
  {https://doi.org/10.1103/PhysRevB.74.161403} {\bibfield  {journal} {\bibinfo
  {journal} {Phys. Rev. B}\ }\textbf {\bibinfo {volume} {74}},\ \bibinfo
  {pages} {161403(R)} (\bibinfo {year} {2006})}\BibitemShut {NoStop}%
\bibitem [{\citenamefont {Oostinga}\ \emph {et~al.}(2008)\citenamefont
  {Oostinga}, \citenamefont {Heersche}, \citenamefont {Liu}, \citenamefont
  {Morpurgo},\ and\ \citenamefont {Vandersypen}}]{Oostinga2008BLG}%
  \BibitemOpen
  \bibfield  {author} {\bibinfo {author} {\bibfnamefont {J.~B.}\ \bibnamefont
  {Oostinga}}, \bibinfo {author} {\bibfnamefont {H.~B.}\ \bibnamefont
  {Heersche}}, \bibinfo {author} {\bibfnamefont {X.}~\bibnamefont {Liu}},
  \bibinfo {author} {\bibfnamefont {A.~F.}\ \bibnamefont {Morpurgo}},\ and\
  \bibinfo {author} {\bibfnamefont {L.~M.}\ \bibnamefont {Vandersypen}},\
  }\bibfield  {title} {\bibinfo {title} {Gate-induced insulating state in
  bilayer graphene devices},\ }\href {https://doi.org/10.1038/nmat2082}
  {\bibfield  {journal} {\bibinfo  {journal} {Nat. Mat.}\ }\textbf {\bibinfo
  {volume} {7}},\ \bibinfo {pages} {151–157} (\bibinfo {year}
  {2008})}\BibitemShut {NoStop}%
\bibitem [{\citenamefont {Ihn}(2010)}]{Thomasbook}%
  \BibitemOpen
  \bibfield  {author} {\bibinfo {author} {\bibfnamefont {T.}~\bibnamefont
  {Ihn}},\ }\href@noop {} {\emph {\bibinfo {title} {Semiconductor
  Nanostructures: Quantum states and electronic transport}}}\ (\bibinfo
  {publisher} {Oxford University Press},\ \bibinfo {year} {2010})\BibitemShut
  {NoStop}%
\bibitem [{\citenamefont {Banszerus}\ \emph {et~al.}(2020)\citenamefont
  {Banszerus}, \citenamefont {Rothstein}, \citenamefont {Fabian}, \citenamefont
  {Möller}, \citenamefont {Icking}, \citenamefont {Trellenkamp}, \citenamefont
  {Lentz}, \citenamefont {Neumaier}, \citenamefont {Watanabe}, \citenamefont
  {Taniguchi}, \citenamefont {Libisch}, \citenamefont {Volk},\ and\
  \citenamefont {Stampfer}}]{aachenehcrossoover}%
  \BibitemOpen
  \bibfield  {author} {\bibinfo {author} {\bibfnamefont {L.}~\bibnamefont
  {Banszerus}}, \bibinfo {author} {\bibfnamefont {A.}~\bibnamefont
  {Rothstein}}, \bibinfo {author} {\bibfnamefont {T.}~\bibnamefont {Fabian}},
  \bibinfo {author} {\bibfnamefont {S.}~\bibnamefont {Möller}}, \bibinfo
  {author} {\bibfnamefont {E.}~\bibnamefont {Icking}}, \bibinfo {author}
  {\bibfnamefont {S.}~\bibnamefont {Trellenkamp}}, \bibinfo {author}
  {\bibfnamefont {F.}~\bibnamefont {Lentz}}, \bibinfo {author} {\bibfnamefont
  {D.}~\bibnamefont {Neumaier}}, \bibinfo {author} {\bibfnamefont
  {K.}~\bibnamefont {Watanabe}}, \bibinfo {author} {\bibfnamefont
  {T.}~\bibnamefont {Taniguchi}}, \bibinfo {author} {\bibfnamefont
  {F.}~\bibnamefont {Libisch}}, \bibinfo {author} {\bibfnamefont
  {C.}~\bibnamefont {Volk}},\ and\ \bibinfo {author} {\bibfnamefont
  {C.}~\bibnamefont {Stampfer}},\ }\bibfield  {title} {\bibinfo {title}
  {Electron-hole crossover in gate-controlled bilayer graphene quantum dots},\
  }\href@noop {} {\  (\bibinfo {year} {2020})},\ \Eprint
  {https://arxiv.org/abs/2008.02585} {arXiv:2008.02585 [cond-mat.mes-hall]}
  \BibitemShut {NoStop}%
\bibitem [{\citenamefont {Eich}\ \emph
  {et~al.}(2018{\natexlab{b}})\citenamefont {Eich}, \citenamefont {Pisoni},
  \citenamefont {Pally}, \citenamefont {Overweg}, \citenamefont {Kurzmann},
  \citenamefont {Lee}, \citenamefont {Rickhaus}, \citenamefont {Watanabe},
  \citenamefont {Taniguchi}, \citenamefont {Ensslin} \emph
  {et~al.}}]{mariusnanolett}%
  \BibitemOpen
  \bibfield  {author} {\bibinfo {author} {\bibfnamefont {M.}~\bibnamefont
  {Eich}}, \bibinfo {author} {\bibfnamefont {R.}~\bibnamefont {Pisoni}},
  \bibinfo {author} {\bibfnamefont {A.}~\bibnamefont {Pally}}, \bibinfo
  {author} {\bibfnamefont {H.}~\bibnamefont {Overweg}}, \bibinfo {author}
  {\bibfnamefont {A.}~\bibnamefont {Kurzmann}}, \bibinfo {author}
  {\bibfnamefont {Y.}~\bibnamefont {Lee}}, \bibinfo {author} {\bibfnamefont
  {P.}~\bibnamefont {Rickhaus}}, \bibinfo {author} {\bibfnamefont
  {K.}~\bibnamefont {Watanabe}}, \bibinfo {author} {\bibfnamefont
  {T.}~\bibnamefont {Taniguchi}}, \bibinfo {author} {\bibfnamefont
  {K.}~\bibnamefont {Ensslin}}, \emph {et~al.},\ }\bibfield  {title} {\bibinfo
  {title} {Coupled quantum dots in bilayer graphene},\ }\href
  {https://doi.org/10.1021/acs.nanolett.8b01859} {\bibfield  {journal}
  {\bibinfo  {journal} {Nano Lett.}\ }\textbf {\bibinfo {volume} {18}},\
  \bibinfo {pages} {5042} (\bibinfo {year} {2018}{\natexlab{b}})}\BibitemShut
  {NoStop}%
\bibitem [{\citenamefont {Overweg}\ \emph
  {et~al.}(2018{\natexlab{a}})\citenamefont {Overweg}, \citenamefont
  {Eggimann}, \citenamefont {Chen}, \citenamefont {Slizovskiy}, \citenamefont
  {Eich}, \citenamefont {Pisoni}, \citenamefont {Lee}, \citenamefont
  {Rickhaus}, \citenamefont {Watanabe}, \citenamefont {Taniguchi} \emph
  {et~al.}}]{Hiske2018electrostatically}%
  \BibitemOpen
  \bibfield  {author} {\bibinfo {author} {\bibfnamefont {H.}~\bibnamefont
  {Overweg}}, \bibinfo {author} {\bibfnamefont {H.}~\bibnamefont {Eggimann}},
  \bibinfo {author} {\bibfnamefont {X.}~\bibnamefont {Chen}}, \bibinfo {author}
  {\bibfnamefont {S.}~\bibnamefont {Slizovskiy}}, \bibinfo {author}
  {\bibfnamefont {M.}~\bibnamefont {Eich}}, \bibinfo {author} {\bibfnamefont
  {R.}~\bibnamefont {Pisoni}}, \bibinfo {author} {\bibfnamefont
  {Y.}~\bibnamefont {Lee}}, \bibinfo {author} {\bibfnamefont {P.}~\bibnamefont
  {Rickhaus}}, \bibinfo {author} {\bibfnamefont {K.}~\bibnamefont {Watanabe}},
  \bibinfo {author} {\bibfnamefont {T.}~\bibnamefont {Taniguchi}}, \emph
  {et~al.},\ }\bibfield  {title} {\bibinfo {title} {Electrostatically induced
  quantum point contacts in bilayer graphene},\ }\href
  {https://doi.org/10.1021/acs.nanolett.7b04666} {\bibfield  {journal}
  {\bibinfo  {journal} {Nano Lett.}\ }\textbf {\bibinfo {volume} {18}},\
  \bibinfo {pages} {553} (\bibinfo {year} {2018}{\natexlab{a}})}\BibitemShut
  {NoStop}%
\bibitem [{\citenamefont {Wang}\ \emph {et~al.}(2013)\citenamefont {Wang},
  \citenamefont {Meric}, \citenamefont {Huang}, \citenamefont {Gao},
  \citenamefont {Gao}, \citenamefont {Tran}, \citenamefont {Taniguchi},
  \citenamefont {Watanabe}, \citenamefont {Campos}, \citenamefont {Muller}
  \emph {et~al.}}]{wang2013drytransferedge}%
  \BibitemOpen
  \bibfield  {author} {\bibinfo {author} {\bibfnamefont {L.}~\bibnamefont
  {Wang}}, \bibinfo {author} {\bibfnamefont {I.}~\bibnamefont {Meric}},
  \bibinfo {author} {\bibfnamefont {P.}~\bibnamefont {Huang}}, \bibinfo
  {author} {\bibfnamefont {Q.}~\bibnamefont {Gao}}, \bibinfo {author}
  {\bibfnamefont {Y.}~\bibnamefont {Gao}}, \bibinfo {author} {\bibfnamefont
  {H.}~\bibnamefont {Tran}}, \bibinfo {author} {\bibfnamefont {T.}~\bibnamefont
  {Taniguchi}}, \bibinfo {author} {\bibfnamefont {K.}~\bibnamefont {Watanabe}},
  \bibinfo {author} {\bibfnamefont {L.}~\bibnamefont {Campos}}, \bibinfo
  {author} {\bibfnamefont {D.}~\bibnamefont {Muller}}, \emph {et~al.},\
  }\bibfield  {title} {\bibinfo {title} {One-dimensional electrical contact to
  a two-dimensional material},\ }\href
  {https://doi.org/10.1126/science.1244358} {\bibfield  {journal} {\bibinfo
  {journal} {Science}\ }\textbf {\bibinfo {volume} {342}},\ \bibinfo {pages}
  {614} (\bibinfo {year} {2013})}\BibitemShut {NoStop}%
\bibitem [{\citenamefont {Overweg}\ \emph
  {et~al.}(2018{\natexlab{b}})\citenamefont {Overweg}, \citenamefont {Knothe},
  \citenamefont {Fabian}, \citenamefont {Linhart}, \citenamefont {Rickhaus},
  \citenamefont {Wernli}, \citenamefont {Watanabe}, \citenamefont {Taniguchi},
  \citenamefont {S{\'a}nchez}, \citenamefont {Burgd{\"o}rfer}, \citenamefont
  {Libisch}, \citenamefont {Fal'ko}, \citenamefont {Ensslin},\ and\
  \citenamefont {Ihn}}]{Overweg2018a}%
  \BibitemOpen
  \bibfield  {author} {\bibinfo {author} {\bibfnamefont {H.}~\bibnamefont
  {Overweg}}, \bibinfo {author} {\bibfnamefont {A.}~\bibnamefont {Knothe}},
  \bibinfo {author} {\bibfnamefont {T.}~\bibnamefont {Fabian}}, \bibinfo
  {author} {\bibfnamefont {L.}~\bibnamefont {Linhart}}, \bibinfo {author}
  {\bibfnamefont {P.}~\bibnamefont {Rickhaus}}, \bibinfo {author}
  {\bibfnamefont {L.}~\bibnamefont {Wernli}}, \bibinfo {author} {\bibfnamefont
  {K.}~\bibnamefont {Watanabe}}, \bibinfo {author} {\bibfnamefont
  {T.}~\bibnamefont {Taniguchi}}, \bibinfo {author} {\bibfnamefont
  {D.}~\bibnamefont {S{\'a}nchez}}, \bibinfo {author} {\bibfnamefont
  {J.}~\bibnamefont {Burgd{\"o}rfer}}, \bibinfo {author} {\bibfnamefont
  {F.}~\bibnamefont {Libisch}}, \bibinfo {author} {\bibfnamefont {V.~I.}\
  \bibnamefont {Fal'ko}}, \bibinfo {author} {\bibfnamefont {K.}~\bibnamefont
  {Ensslin}},\ and\ \bibinfo {author} {\bibfnamefont {T.}~\bibnamefont {Ihn}},\
  }\bibfield  {title} {\bibinfo {title} {Topologically {{Nontrivial Valley
  States}} in {{Bilayer Graphene Quantum Point Contacts}}},\ }\href
  {https://doi.org/10.1103/PhysRevLett.121.257702} {\bibfield  {journal}
  {\bibinfo  {journal} {Physical Review Letters}\ }\textbf {\bibinfo {volume}
  {121}},\ \bibinfo {pages} {257702} (\bibinfo {year}
  {2018}{\natexlab{b}})}\BibitemShut {NoStop}%
\bibitem [{\citenamefont {McCann}\ \emph {et~al.}(2007)\citenamefont {McCann},
  \citenamefont {Abergel},\ and\ \citenamefont {Fal'ko}}]{McCann2007}%
  \BibitemOpen
  \bibfield  {author} {\bibinfo {author} {\bibfnamefont {E.}~\bibnamefont
  {McCann}}, \bibinfo {author} {\bibfnamefont {D.~S.}\ \bibnamefont
  {Abergel}},\ and\ \bibinfo {author} {\bibfnamefont {V.~I.}\ \bibnamefont
  {Fal'ko}},\ }\bibfield  {title} {\bibinfo {title} {The low energy electronic
  band structure of bilayer graphene},\ }\href
  {https://doi.org/10.1140/epjst/e2007-00229-1} {\bibfield  {journal} {\bibinfo
   {journal} {The European Physical Journal Special Topics}\ }\textbf {\bibinfo
  {volume} {148}},\ \bibinfo {pages} {91} (\bibinfo {year} {2007})}\BibitemShut
  {NoStop}%
\bibitem [{\citenamefont {McCann}\ and\ \citenamefont
  {Koshino}(2013)}]{McCann2013}%
  \BibitemOpen
  \bibfield  {author} {\bibinfo {author} {\bibfnamefont {E.}~\bibnamefont
  {McCann}}\ and\ \bibinfo {author} {\bibfnamefont {M.}~\bibnamefont
  {Koshino}},\ }\bibfield  {title} {\bibinfo {title} {The electronic properties
  of bilayer graphene},\ }\href {https://doi.org/10.1088/0034-4885/76/5/056503}
  {\bibfield  {journal} {\bibinfo  {journal} {Reports on Progress in Physics}\
  }\textbf {\bibinfo {volume} {76}},\ \bibinfo {pages} {056503} (\bibinfo
  {year} {2013})}\BibitemShut {NoStop}%
\bibitem [{\citenamefont {Varlet}\ \emph {et~al.}(2014)\citenamefont {Varlet},
  \citenamefont {Bischoff}, \citenamefont {Simonet}, \citenamefont {Watanabe},
  \citenamefont {Taniguchi}, \citenamefont {Ihn}, \citenamefont {Ensslin},
  \citenamefont {{Mucha-Kruczy{\'n}ski}},\ and\ \citenamefont
  {Fal'ko}}]{Varlet2014}%
  \BibitemOpen
  \bibfield  {author} {\bibinfo {author} {\bibfnamefont {A.}~\bibnamefont
  {Varlet}}, \bibinfo {author} {\bibfnamefont {D.}~\bibnamefont {Bischoff}},
  \bibinfo {author} {\bibfnamefont {P.}~\bibnamefont {Simonet}}, \bibinfo
  {author} {\bibfnamefont {K.}~\bibnamefont {Watanabe}}, \bibinfo {author}
  {\bibfnamefont {T.}~\bibnamefont {Taniguchi}}, \bibinfo {author}
  {\bibfnamefont {T.}~\bibnamefont {Ihn}}, \bibinfo {author} {\bibfnamefont
  {K.}~\bibnamefont {Ensslin}}, \bibinfo {author} {\bibfnamefont
  {M.}~\bibnamefont {{Mucha-Kruczy{\'n}ski}}},\ and\ \bibinfo {author}
  {\bibfnamefont {V.~I.}\ \bibnamefont {Fal'ko}},\ }\bibfield  {title}
  {\bibinfo {title} {Anomalous {{Sequence}} of {{Quantum Hall Liquids
  Revealing}} a {{Tunable Lifshitz Transition}} in {{Bilayer Graphene}}},\
  }\href {https://doi.org/10.1103/PhysRevLett.113.116602} {\bibfield  {journal}
  {\bibinfo  {journal} {Physical Review Letters}\ }\textbf {\bibinfo {volume}
  {113}},\ \bibinfo {pages} {116602} (\bibinfo {year} {2014})}\BibitemShut
  {NoStop}%
\bibitem [{\citenamefont {Varlet}\ \emph {et~al.}(2015)\citenamefont {Varlet},
  \citenamefont {{Mucha-Kruczy{\'n}ski}}, \citenamefont {Bischoff},
  \citenamefont {Simonet}, \citenamefont {Taniguchi}, \citenamefont {Watanabe},
  \citenamefont {Fal'ko}, \citenamefont {Ihn},\ and\ \citenamefont
  {Ensslin}}]{Varlet2015}%
  \BibitemOpen
  \bibfield  {author} {\bibinfo {author} {\bibfnamefont {A.}~\bibnamefont
  {Varlet}}, \bibinfo {author} {\bibfnamefont {M.}~\bibnamefont
  {{Mucha-Kruczy{\'n}ski}}}, \bibinfo {author} {\bibfnamefont {D.}~\bibnamefont
  {Bischoff}}, \bibinfo {author} {\bibfnamefont {P.}~\bibnamefont {Simonet}},
  \bibinfo {author} {\bibfnamefont {T.}~\bibnamefont {Taniguchi}}, \bibinfo
  {author} {\bibfnamefont {K.}~\bibnamefont {Watanabe}}, \bibinfo {author}
  {\bibfnamefont {V.}~\bibnamefont {Fal'ko}}, \bibinfo {author} {\bibfnamefont
  {T.}~\bibnamefont {Ihn}},\ and\ \bibinfo {author} {\bibfnamefont
  {K.}~\bibnamefont {Ensslin}},\ }\bibfield  {title} {\bibinfo {title} {Tunable
  {{Fermi}} surface topology and {{Lifshitz}} transition in bilayer graphene},\
  }\href {https://doi.org/10.1016/j.synthmet.2015.07.006} {\bibfield  {journal}
  {\bibinfo  {journal} {Synthetic Metals}\ }\bibinfo {series} {Reviews of
  {{Current Advances}} in {{Graphene Science}} and {{Technology}}},\ \textbf
  {\bibinfo {volume} {210}},\ \bibinfo {pages} {19} (\bibinfo {year}
  {2015})}\BibitemShut {NoStop}%
\bibitem [{\citenamefont {Xiao}\ \emph {et~al.}(2010)\citenamefont {Xiao},
  \citenamefont {Chang},\ and\ \citenamefont {Niu}}]{Xiao2010}%
  \BibitemOpen
  \bibfield  {author} {\bibinfo {author} {\bibfnamefont {D.}~\bibnamefont
  {Xiao}}, \bibinfo {author} {\bibfnamefont {M.-C.}\ \bibnamefont {Chang}},\
  and\ \bibinfo {author} {\bibfnamefont {Q.}~\bibnamefont {Niu}},\ }\bibfield
  {title} {\bibinfo {title} {Berry phase effects on electronic properties},\
  }\href {https://doi.org/10.1103/RevModPhys.82.1959} {\bibfield  {journal}
  {\bibinfo  {journal} {Reviews of Modern Physics}\ }\textbf {\bibinfo {volume}
  {82}},\ \bibinfo {pages} {1959} (\bibinfo {year} {2010})}\BibitemShut
  {NoStop}%
\bibitem [{\citenamefont {Chang}\ and\ \citenamefont {Niu}(1996)}]{Chang1996a}%
  \BibitemOpen
  \bibfield  {author} {\bibinfo {author} {\bibfnamefont {M.-C.}\ \bibnamefont
  {Chang}}\ and\ \bibinfo {author} {\bibfnamefont {Q.}~\bibnamefont {Niu}},\
  }\bibfield  {title} {\bibinfo {title} {Berry phase, hyperorbits, and the
  {{Hofstadter}} spectrum: {{Semiclassical}} dynamics in magnetic {{Bloch}}
  bands},\ }\href {https://doi.org/10.1103/PhysRevB.53.7010} {\bibfield
  {journal} {\bibinfo  {journal} {Physical Review B}\ }\textbf {\bibinfo
  {volume} {53}},\ \bibinfo {pages} {7010} (\bibinfo {year}
  {1996})}\BibitemShut {NoStop}%
\end{thebibliography}
\end{document}